\documentclass[twocolumn]{autart}
\usepackage{natbib}            
\usepackage{graphicx}          

\usepackage{array}
\usepackage{latexsym,theorem}
\usepackage{amssymb,amsfonts,amsmath}
\usepackage{multirow}
\usepackage{xcolor}
\usepackage{enumitem}

\newcommand*{\red}{\textcolor{black}}
\newcommand{\smallsub}[1]{\! {\scriptscriptstyle \mathcal{#1}} }
\newcommand{\subsuper}[3]{{{#1}_{#2}}\!^{#3}}
{\theorembodyfont{\slshape}\newtheorem{theorem}{Theorem}}
{\theorembodyfont{\slshape}\newtheorem{proposition}[theorem]{Proposition}}
{\theorembodyfont{\slshape}}
{\theorembodyfont{\slshape}\newtheorem{corollary}[theorem]{Corollary}}
{\theorembodyfont{\slshape}\newtheorem{definition}{Definition}}
{\theorembodyfont{\upshape}}
{\theorembodyfont{\upshape}\newtheorem{remark}{Remark}}

\newcommand{\mY}{\mathcal{Y}}
\newcommand{\mX}{\mathcal{X}}
\newcommand{\mD}{\mathcal{D}}

\newcommand{\mG}{\mathcal{G}}
\newcommand{\mE}{\mathcal{E}}
\newcommand{\mV}{\mathcal{V}}

\newcommand{\mT}{\mathcal{T}}

\newcommand{\smB}{\smallsub{B}}

\newcommand{\smF}{\smallsub{F}}

\newcommand{\smD}{\smallsub{D}}
\newcommand{\smY}{\smallsub{Y}}

\newcommand{\smE}{\smallsub{T}}
\newcommand{\smT}{\smallsub{T}}

\newcommand{\Eb}{\bar{\mathbb{E}}}

\newcommand{\beq}{\begin{equation}}
\newcommand{\eeq}{\end{equation}}
\newcommand{\beqr}{\begin{eqnarray}}
\newcommand{\eeqr}{\end{eqnarray}}

\newcommand{\R}{\mathbb{R}}
\newcommand{\eio}{e^{i\omega}}
\newcommand{\M}{\mathcal{M}}

\begin{document}

\begin{frontmatter}

\title{Data-informativity conditions for structured linear systems with implications for dynamic networks}

\thanks[footnoteinfo]{Funded by the European Union. Views and opinions expressed are however those of the author(s) only and do not necessarily reflect those of the European Union or the European Research Council. Neither the European Union nor the European Research Council can be held responsible for them.}
\thanks[footnoteinfo]{This work was supported by VINNOVA Competence Center AdBIOPRO 2016-05181, and by the
Swedish Research Council through contracts [2019-04956, 2025-04833].}

\thanks[date]{\mbox{  }\ \ Original version 20 November 2024; revised 29 July 2025 and 7 January 2026. Final version 17 April 2026.}

\author[TUe]{Paul M.J. Van den Hof}\ead{p.m.j.vandenhof@tue.nl},
\author[DCSC]{Shengling Shi},
\author[VITO]{Stefanie J.M. Fonken},
\author[ASML]{Karthik R. Ramaswamy},
\author[KTH]{H\aa kan Hjalmarsson} and
\author[CAL]{Arne G. Dankers}
\address[TUe]{Control Systems Group, Department of Electrical Engineering, Eindhoven University of Technology, The Netherlands.}
\address[DCSC]{Delft Center for Systems and Control, Delft University of Technology, The Netherlands.}
\address[VITO]{Unit Water and Energy Transition, Flemish Institute for Technological Research (VITO), Mol, Belgium, and EnergyVille, Genk, Belgium.}
\address[ASML]{ASML, Veldhoven, The Netherlands.}
\address[KTH]{Department of Decision and Control Systems, Royal Institute of Technology, Sweden.}
\address[CAL]{Department of Electrical and Software Engineering, University of Calgary, Canada.}

\begin{abstract}                
When estimating a single subsystem (module) in a linear dynamic network with a prediction error method, a data-informativity condition needs to be satisfied for arriving at a consistent module estimate. This concerns a condition on input signals in the constructed, possibly MIMO (multiple input multiple output) predictor model being persistently exciting, which is typically guaranteed if the input spectrum is positive definite for a sufficient number of frequencies. Generically, the condition can be formulated as a path-based condition on the graph of the network model. The current condition has two elements of possible conservatism: (a) rather than focussing on the full MIMO model, one would like to be able to focus on consistently estimating the target module only, and (b) structural information, such as structural zero elements in the interconnection structure or known subsystems, should be taken into account. In this paper relaxed conditions for data-informativity are derived addressing these two issues, leading to relaxed path-based conditions on the network graph. This leads to experimental conditions that are less strict, i.e. require a smaller number of external excitation signals. Additionally, the new expressions for data-informativity in identification are shown to be closely related to earlier derived conditions for (generic) single module identifiability.
\end{abstract}


\end{frontmatter}

\section{Introduction}
\label{sec:intro}
While the dynamic systems that are currently addressed in engineering and science, become more and more complex, large-scale and interconnected, the attention for modelling, monitoring and control aspects of interconnected systems, like e.g. dynamic networks, is growing. For the resulting control problems this has led to the development of decentralized, distributed and multi-agent type of control systems. On the modelling side, the development of analysis tools for systems over graphs has to be mentioned, and we have seen the development of a field called data-driven modeling (or system identification) in dynamic networks. This paper is situated in this latter area, that will be overviewed next.\\
Identification of/in dynamic networks has been developed in a framework where dynamic systems are interconnected in a particular interconnection structure (topology), including cycles, that may be known a priori. In such a setting several different problems have been considered. Following the early contributions on uniqueness of network representations \cite{Goncalves&Warnick:08} and topology estimation \cite{Materassi&Innocenti:10,Materassi&Salapaka:12}, a framework for network identification in a prediction error setting is introduced in \cite{VandenHof&etal_Autom:13}, while subspace methods for network identification have been developed in \cite{Haber&Verhaegen_TAC:14,Verhaegen&etal:22}. Alternatively, there are network identification approaches based on conditional independence and dynamic factor models, see e.g. \cite{Zorzi&Chiuso:17,Veedu&Doddi&Salapaka_TAC:22}. \\
The vast literature on network identification problems in a prediction error setting, ranges from the problem of identifying the full network for a given topology \cite{Yuan:11,Weerts&etal_Autom:18_reducedrank,Fonken&etal_Autom:22}, and the analysis and synthesis of identifiability conditions \cite{Weerts&etal_Autom:18_identifiability,Hendrickx&Gevers&Bazanella_TAC:19,vanWaarde&etal_TAC:20a,Legat&Hendrickx_CDC:20,Cheng&etal_TAC:22,Shi&etal_TAC:23,Mapurunga&etal:24},
to the problem of identifying the dynamics of a single module/link/subsystem in the network \cite{VandenHof&etal_Autom:13,Dankers&etal_TAC:16,Linder&Enqvist_ijc:17,Gevers&etal_sysid:18,Everitt&Bottegal&Hjalmarsson_Autom:18,Materassi&Salapaka:20,Ramaswamy&VandenHof_TAC:21}, which includes the problem of determining where to measure and where to excite the network, and the related data-informativity questions \cite{Bombois&etal_Autom:23}.\\
In this paper we will focus on this latter single module identification problem, in a setting where some dynamic modules in the network may be known a priori, while a single target module will need to be estimated consistently. We focus on the so-called {\it local direct method} \citep{Ramaswamy&VandenHof_TAC:21} that is known to be able to exploit both measured external excitation signals and unmeasured disturbance signals for estimation.
It has been shown \citep{Ramaswamy&VandenHof_TAC:21} that it can be necessary to build a predictor model that has multiple inputs and multiple outputs (MIMO), in order to arrive at consistent estimates of a single target module. This is due to the necessity to deal with {\it confounding variables}, i.e., correlated disturbances between inputs and outputs of an estimation problem, a situation that can typically occur in dynamic networks. The resulting MIMO predictor model then will need to satisfy {\it data-informativity} conditions in order to arrive at consistent module estimates. The objective of this paper is to formulate {\it path-based conditions}, dependent on the a priori known graph of the network only, that guarantee data-informativity, and therefore will provide the conditions for consistent module estimates. These path-based conditions will refer to the locations of external signals (measured excitation signals and unmeasured disturbance signals) that enter the network.\\
For specifying the role of a data-informativity condition, we need to go back to their origin, developed for classical (open-loop type of) identification problems in a prediction error setting. It is most often formulated as a condition on the input spectrum matrix being positive definite at a sufficient number of frequencies \citep{Soderstrom&Stoica:89}. The condition is typically applied to situations where the inputs are general quasi-stationary signals with continuous spectral densities. For other types of inputs, as e.g. periodic signals, the concerned condition is conservative and replaced by conditions on the selection of a sufficient number of frequency components for each input and orthogonality of the signals over the different inputs  \citep{Pintelon&Schoukens:12}. For particular black box model structures with finite order model sets, the spectrum condition can be further detailed, as shown e.g., in \cite{Gevers&Bazanella&Miskovic_CDC:08}. The available theory for data informativity analysis seems to be mainly directed towards the situation of unstructured MIMO black box models, meaning that all entries in the MIMO model are parametrized. A relevant question that pops up is then: what are the consequences for the conditions for data-informativity if particular entries in the MIMO model are structurally zero, or non-zero but a priori known and therefore not necessarily parametrized?  This latter situation is typically present in a dynamic network situation.\\
In the current literature on dynamic network identification, the data-informativity conditions so far have been taken from a general non-structured MIMO estimation problem, i.e., requiring $\Phi_{w_{\smD}}(\omega) \succ 0$ for a sufficient number of frequencies $\omega$, with $w_{\smD}$ the set of inputs nodes in the predictor model. For a MISO setting, the data-informativity conditions have also been considered in \cite{Bombois&etal_Autom:23} in the form of rank conditions on particular system-dependent matrices. For a generic satisfaction of the data-informativity conditions, very effective use has been made of a result of \cite{vanderWoude:91} and \cite{Hendrickx&Gevers&Bazanella_TAC:19} where it is shown that rank conditions on transfer functions can generically be verified by path-based conditions on the graph of the network. This has led to the integration of path-based conditions in the selection of inputs and outputs of the predictor model to warrant that data-informativity conditions can be satisfied generically \citep{VandenHof&Ramaswamy_CDC:20,VandenHof&etal_IFAC:23}.\\
In all these approaches, the data-informativity conditions for black box, fully parametrized MIMO models have been exploited. However, in the considered network problem, structural information on the multivarible predictor model might be readily available (e.g. its topology, indicating which nodes are connected with which other nodes), and rather than consistently estimating the full MIMO predictor model, we can focus on consistently estimating the single target module only.\\
The question that we address in this paper is therefore: in the setting of the mentioned local direct method \cite{Ramaswamy&VandenHof_TAC:21}, can we exploit structural information (topology) of the network, in the form of a priori known entries in the predictor model, to relax the data-informativity conditions and to focus these conditions on estimating a single target module only. In comparison with the current data-informativity literature, these are two innovative steps to be taken: including the structural information, and addressing a single target module only.\\
We will analyze the situation by first addressing the structural information in a standard (open-loop type of) estimation problem in Sections \ref{sec:model}-\ref{sec:di4ss}, including a presentation of the modeling setup and the notion of data-informativity. After presenting an example in Section \ref{sec:exam}, we will then formulate the new results in terms of path-based conditions on the network graph in Section \ref{sec:path}. With these new results, we can address the single module identification problem in dynamic networks in Section \ref{sec:dynnet}, with some examples collected in Section \ref{sec:ex2}. The proofs of the most important results are collected in an Appendix. We start by presenting a very brief example.

\section{A motivating example}
\label{sec:achiev}
\begin{figure}[htb]
\centerline{\includegraphics[scale=0.45]{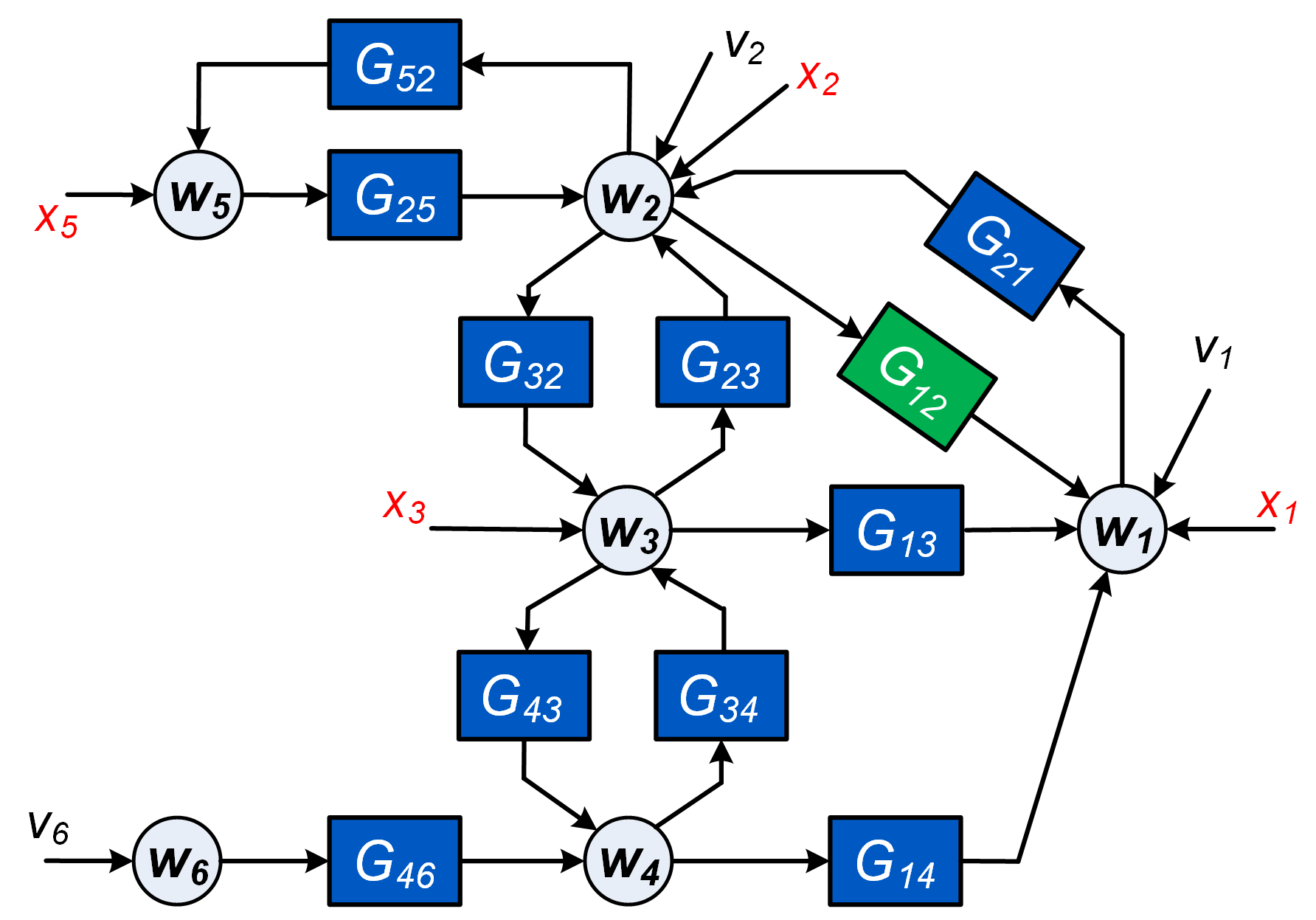}}
	\caption{Six node example with target module $G_{12}(q)$ and correlated disturbances $v_1, v_2$.}
	\label{fig6Anode}
\end{figure}

In order to give a flavor of the type of results that will be achieved in this paper, we present an example network provided in Figure \ref{fig6Anode}, where linear time-invariant systems $G_{jk}$ are interconnected in a network structure, and the node variables $w_{\ell}$ are the output signals of the corresponding summation points. $v_k$ are non-measured stationary stochastic disturbance signals, $v_1$ and $v_2$ being mutually correlated, and $x_i$ are possible locations for adding external signals. Our objective is to identify the target module $G_{12}$ from data $w_{\ell}$, $\ell =1,\cdots 6$, and possibly present measured $x_i$. In the current theory of the local direct method, a consistent estimation of $G_{12}$, can be achieved if the four external signals $x_1$, $x_2$, $x_3$ and $x_5$ are all present, with $x_1$ and $x_2$ being measured.  Our new results, derived in this paper, will show that the presence of signals $x_3$ and $x_5$ only, with $x_5$ being measured, will be sufficient, thereby considerably reducing the experimental burden. This example will be treated in full detail in Section \ref{sec:ex22}.

\section{The modeling setup}
\label{sec:model}
For properly addressing the data-informativity problem, we start by considering a multivariable (open-loop) data-generating
dynamic system described by
\beq
\label{eq:1}
 y(t) = G^0(q)u(t) + H^0(q) e(t)
\eeq
where output $y(t) \in \R^p$, input $u(t) \in \R^m$, the rational transfer function $G^0(\cdot) \in \R^{p\times m}(\cdot)$, $q$ the shift operator $qu(t) = u(t+1)$, $e(t) \in \R^p$ a multivariate white-noise process, and $H^0(\cdot) \in \R^{p\times p}(\cdot)$ a rational noise model, being monic, i.e. $\lim_{z\rightarrow\infty} H(z) = I$, stable and stably invertible. For simplicity we will assume that $G^0(q)$ is stable too.\\
The data-generating system is modeled through a parame\-trized model $(G(q,\theta), H(q,\theta))$, with parameters $\theta \in \Theta \subset \R^n$. The model's one-step ahead predictor can be defined as $\hat y(t|t-1;\theta) := \Eb [y(t)|y^{t-1},u^t]$ where $u^t$ refers to the past of signal $u$ up to time instant $t$, and $\Eb$ is the generalized expectation operator $\Eb := \lim_{N\rightarrow\infty} \frac{1}{N} \sum_{t=1}^N \mathbb{E}$ for quasi-stationary signals (\cite{Ljung:99}).

This leads to the classical expression for the one-step-ahead predictor:
\beq
\label{eqyp}
\hat y(t|t-1;\theta) = \underbrace{\begin{bmatrix} (I-H(q,\theta)^{-1}) & H(q,\theta)^{-1}G(q,\theta)\end{bmatrix}}_{W(q,\theta)} \underbrace{\begin{bmatrix} y(t) \\ u(t) \end{bmatrix}}_{z(t)}
\eeq
with predictor filter $W(q,\theta)$.\\
In line with the corresponding definitions in the prediction error literature (\cite{Ljung:99}, Definition 8.1), we can now define the notion of data-informativity for the related predictor model.
\begin{definition} \label{def1}
Consider a set of signals contained in $z$ and a predictor model
$\hat y(t|t-1;\theta) = W(q,\theta)z(t)$,
 as in (\ref{eqyp}) for a parametrized set of models
$\mathcal{M} := (G(q,\theta),H(q,\theta))_{\theta\in\Theta}$.
Then a quasi-stationary data set $Z^{\infty}:=\{z(t)\}_{t=0,\cdots,\infty}$
is {\it informative enough with respect to the model set $\mathcal{M}$} if, for any two predictor models $W(q,\theta_1)$ and $W(q,\theta_2)$ in the model set, the equation
\beq
\label{eq:wq}
  \Eb[\|(W(q,\theta_1)-W(q,\theta_2))z(t)\|^2] = 0
\eeq
implies that $W(\eio,\theta_1)\equiv W_2(\eio,\theta_2)$ for almost all $\omega$, where $\|\cdot \|$ is the standard Euclidean norm.\hfill $\Box$
\end{definition}

Data-informativity guarantees that one can distinguish between different models in a model set, and consequently, that one can identify unique models on the basis of an experimental data set of infinite length. It is a necessary condition for the consistent estimation of models. If we do not constrain the order of the models in the model set $\mathcal{M}$, but allow $\mathcal{M}$ to consist of all linear time-invariant models of any order, then a typical sufficient condition for data-informativity is the condition that
$\Phi_z(\omega) \succ 0$ for almost all $\omega$.
For open-loop situations where the signals $u$ and $e$ are uncorrelated, this can be further relaxed to the condition
\beq
\label{eq:fu}
        \Phi_u(\omega) \succ 0 \ \ \ \mbox{for almost all } \omega,
\eeq
being referred to as the property of input signal $u$ being persistently exciting (\cite{Ljung:99}).

\section{Data-informativity conditions for consistent estimation}
\label{sec:di}

The data-informativity conditions play a central role in the proof of consistency of estimates of $G^0(q)$ and $H^0(q)$ when applying an (asymptotic) prediction error identification criterion
\beq
\label{eq:ts}
   \theta^* = \arg\min_{\theta} \Eb \varepsilon^T(t,\theta) Q\varepsilon(t,\theta)
\eeq
with the one-step-ahead prediction error
\beqr
\varepsilon(t,\theta) & := & y(t) - \hat y(t|t-1;\theta), \nonumber \\
& = & H(q,\theta)^{-1}[y(t) - G(q,\theta)u(t)], \label{eq:8}
\eeqr
and $Q$ any positive definite weighting matrix. Without loss of generality we will use $Q=I$ in the sequel, while for the consistency analysis we will assume that $(G^0,H^0)\in\M$.

Substituting (\ref{eq:1}) into (\ref{eq:8}) leads to
\beqr
\varepsilon(t,\theta) & = & H(q,\theta)^{-1}[G^0(q)-G(q,\theta)] u(t) + \nonumber \\ & & + H(q,\theta)^{-1}[H^0(q)-H(q,\theta)]e(t) + e(t)
\eeqr
which can alternatively be written as
\beq
\varepsilon(t,\theta) = H(q,\theta)^{-1}\Delta M(q,\theta)\kappa(t) + e(t),
\eeq
with $\Delta M(q,\theta):= [\Delta G(q,\theta)\ \ \Delta H(q,\theta)]$,
$\Delta G(q,\theta) := G^0(q)-G(q,\theta)$, $\Delta H(q,\theta) := H^0(q)-H(q,\theta)$ and $\kappa(t) :=  [u^T(t)\ \ e^T(t)]^T$.
The proof of consistency \citep{Ljung:78} comes down to showing that
\begin{itemize}
\item[a.] $\Delta M(q,\theta)=0$ achieves the global minimum of the cost function in (\ref{eq:ts}), and
\item[b.] this minimum is unique, i.e. it provides a single point estimate $(G(q,\theta^*),H(q,\theta^*))$.
\end{itemize}
If the transfer function $\Delta H(q,\theta)$ is strictly proper, which is guaranteed by the assumption that $H^0(q)$ and $H(q,\theta)$ both are monic, then it is straightforward to show that
\beq
\theta^*  =  \arg\min_{\theta} \Eb \| H(q,\theta)^{-1}  \Delta M(q,\theta) \kappa(t) \|^2. \label{eq:jw}
\eeq
With this expression it is obvious that condition (a) above is satisfied, reaching the minimum value of $0$ of the cost function in (\ref{eq:jw}) for $\Delta M = 0$. For the uniqueness condition (b), we need to formulate a condition under which the following implication holds:
\beq
\{\Eb \| H(q,\theta)^{-1} \Delta M(q,\theta) \kappa(t) \|^2 = 0\} \Longrightarrow
\{\Delta M(q,\theta) = 0\}. \nonumber
\eeq
Since $H(q,\theta)$ is stable and stably invertible, for any quasi-stationary signal $x(t)$ it holds that
\[ \{\Eb \| H(q,\theta)^{-1} x(t) \|^2 = 0\} \Longleftrightarrow \{\Eb \| x(t) \|^2 = 0\} \]
so that we can rewrite the required implication for condition (b) as
\beq
\{\Eb \| \Delta M(q,\theta) \kappa(t)\|^2 = 0\} \Longrightarrow
\{\Delta M(q,\theta)= 0\}. \label{eq:imp}
\eeq
Using Parseval's relation the left hand side can be written as:
\beq
\label{eq:pars}
 \frac{1}{2\pi} \mbox{trace} \{\int_{-\pi}^\pi \Delta M(\eio,\theta) \Phi_{\kappa}(\omega) \Delta M^T(e^{-i\omega},\theta) d\omega\} = 0
\eeq
and this expression implies $\Delta M(q,\theta) =0$ if $\Phi_{\kappa}(\omega) \succ 0$ in a sufficient number of frequencies. If no conditions are imposed on the order of $M(q,\theta)$ then the typically used condition is to require $\Phi_{\kappa}(\omega) \succ 0$ at almost all $\omega$. In the open-loop case where $u$ and $e$ are uncorrelated, this reduces to the persistent excitation condition (\ref{eq:fu}).

\section{Data informativity conditions for structured systems}
\label{sec:di4ss}
\subsection{Data-informativity for a single row / output}
\label{sec:row}
We will now consider the situation that the multivariable system $G^0$ is structured, in the sense that there is prior knowledge on the presence of structural zeros in the transfer matrix $G^0$. In order to analyze this situation, we rewrite the expression (\ref{eq:pars}) as
\beq
\label{eq:cf}
 \frac{1}{2\pi} \sum_{\ell=1}^p \{\int_{-\pi}^\pi \Delta M_{\ell}(\eio,\theta) \Phi_{\kappa}(\omega) \Delta M_{\ell}^T(e^{-i\omega},\theta) d\omega\} = 0
\eeq
with $M_{\ell}(\eio,\theta)$ the $\ell$-th row of $M(\eio,\theta)$.

Additionally we define $\kappa^{[\ell]}$ as the subset of signals in $\kappa$ corresponding to the column numbers of the nonzero entries in the row $M_{\ell}(q,\theta)$. In other words: $\kappa^{[\ell]}$ are those signals $u$ and $e$ that in the parametrized model are inputs to nonzero maps to $y_{\ell}$, with $y_{\ell}$ the $\ell$-th component of output vector $y$. Then the following result can be formulated.
\begin{proposition}
\label{prop1}
Consider the model setup as described in Section \ref{sec:model}. If the rows in the model $M_{\ell}(q,\theta)$ are independently parametrized, then
$M(q,\theta) = M^0(q)$ is the unique minimum of the cost function in (\ref{eq:ts}) if, for $\ell = 1, \cdots p$, $\Phi_{\kappa^{[\ell]}}(\omega) \succ 0$ for almost all $\omega$.
\end{proposition}
{\bf Proof}
Since the cost function (\ref{eq:cf}) is written as a sum of $p$ components, with each of the components $\geq 0$, the implication (\ref{eq:imp}) can now be applied to each component separately. Uniqueness of the minimum is then guaranteed if for each individual component the corresponding spectral density is positive definite.
\hfill $\Box$

By considering $\kappa^{[\ell]}$ rather then $\kappa$, we take account of the structural zero entries in row $M_{\ell}(q,\theta)$. Note that if one of the rows in $M(q,\theta)$ has no structural zero entries, then there is no added value in the above result, since it will reduce to the condition $\Phi_{\kappa}(\omega) \succ 0$ for the non-structured case.

However there is an important second consequence of the handling of structure. In the situation that we are actually only interested in consistent estimation of row $\ell$ in the model $M(q,\theta)$, i.e. related to the particular output $y_{\ell}$, then the following result can be formulated.

\begin{proposition}
\label{prop2}
Consider the model setup as described in Section \ref{sec:model}. If the parameters in row $M_j(q,\theta)$ are independent of the parameters in the other rows of $M(q,\theta)$, then $M_j(q,\theta) = M_j^0(q)$ is unique in the minimum of the cost function in (\ref{eq:ts}) if $\Phi_{\kappa^{[j]}}(\omega) \succ 0$ for almost all $\omega$.
\end{proposition}
{\bf Proof}
Since the cost function (\ref{eq:cf}) is actually a summation over $p$ nonnegative terms, the term $\ell = j$ can be minimized independent of the other terms if there are no shared parameters. Uniqueness of the minimum for the term related to $\ell = j$ is then achieved when the corresponding spectrum related to the considered row of $\Delta M(\eio,\theta)$ is positive definite. \hfill $\Box$

This result provides a data-informativity condition for the consistent estimation of the model terms in $M_j(q,\theta)$ irrespective of possible consistency of the other terms in the model $M(q,\theta)$. As a result it is a less conservative condition than the original (non-structured) result $\Phi_{\kappa}(\omega) \succ 0$.

A (prediction error) identification algorithm that would estimate a model by minimizing a scalar cost function of the prediction error (\ref{eq:8}), would, under the data-informativity condition of Proposition \ref{prop2}, typically lead to a situation where part of the model parameters are uniquely determined and estimated consistently, while other parameters are non-unique. For the concerned optimization algorithms, this requires the appropriate handling of this situation in terms of dealing with singular Jacobian matrices during the optimization. Currently applied optimization algorithm, as e.g. implemented in MATLAB's System Identification Toolbox, can handle this effectively.

One may wonder why it would be interesting, when identifying a MIMO model, to focus on the consistency of only a part of the model. This problem  particularly appears when identifying a single module in a dynamic network, as further elaborated in subsequent sections.

\begin{remark}
While the vector $\kappa^{[j]}$ is composed of the signals related to (the columns of) nonzero terms in the corresponding model $[G(q,\theta)\ H(q,\theta)]_{j*}$, the theory actually holds for signals related to (the columns of) {\it parametrized} terms in the corresponding model. In other words, terms in the model that are known a priori (like controllers in a network), can be excluded from $\kappa^{[j]}$, as they do not serve as inputs to terms that are parametrized and need to be identified.
\end{remark}

\subsection{Including a network structure in the generation of input signals}
\label{sec:ustruc}
The result of Proposition \ref{prop2} can be further worked out if, besides structure in the considered MIMO system, we also consider structure in the generation of the input signal $u$. As a further step towards the consideration of dynamic networks, we will first consider the situation that the input signal $u$ has been generated in a  structured way, including dependencies among the different components in $u$.

\begin{figure}[h]
\centerline{\includegraphics[scale=0.5]{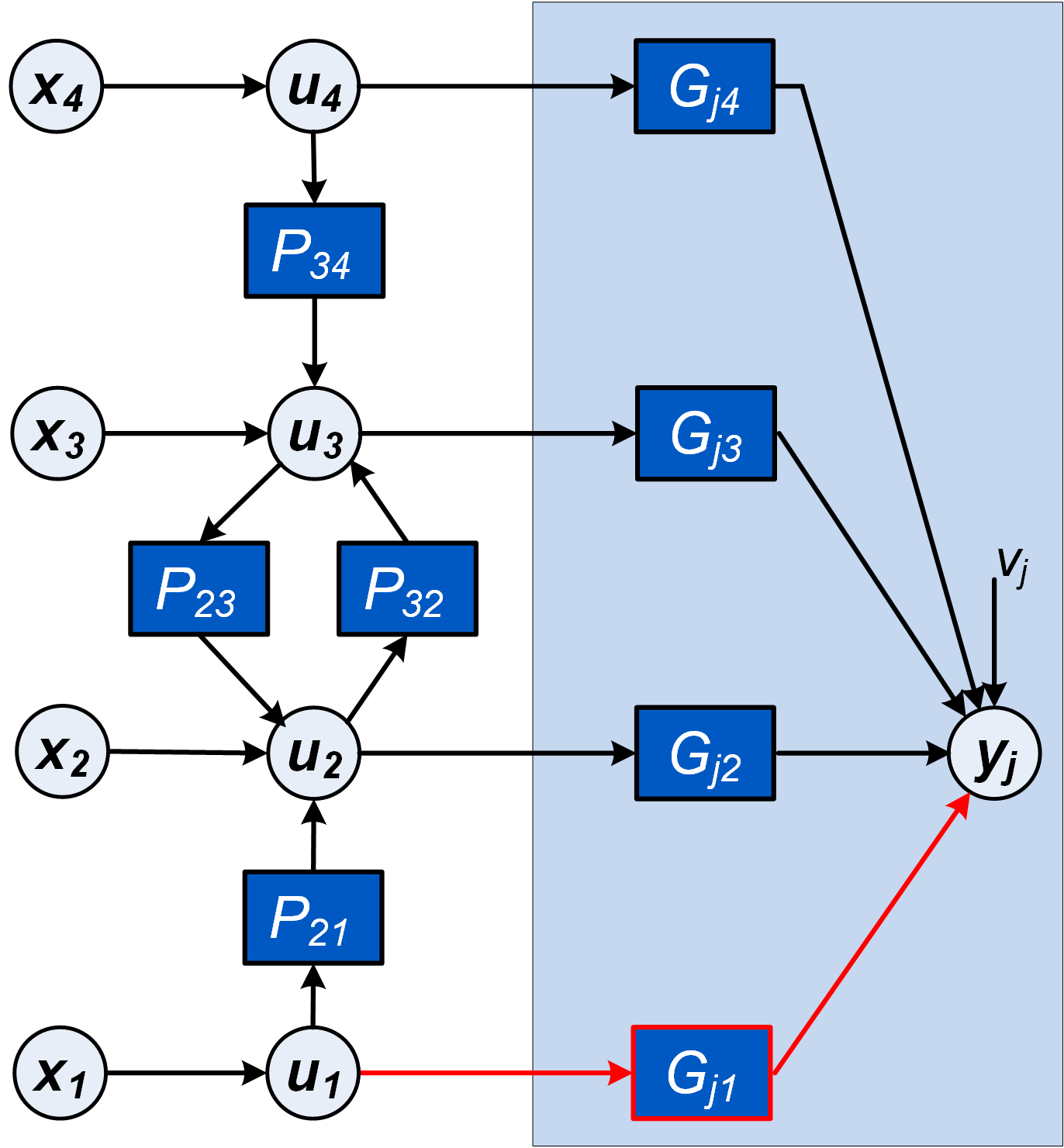}}
	\caption{Five node example with target module $G_{j1}(q)$, and dependence structure among the input components $u$ as in (\ref{equ}), where $v_j$ is the $j$-th component of $H^0(q)e(t)$ in (\ref{eq:1}).}
	\label{fig1}
\end{figure}

In line with the notation used in dynamic networks (see e.g. \cite{VandenHof&etal_Autom:13}), we will consider the input $u$ to be constructed as a multivariable filtered version of an external $n_x$-dimensional vector signal $x$, such that
\beq
\label{equ}
u(t) = P(q) u(t) + Q(q) x(t)
\eeq
with $P$ a $m \times m$ proper and stable transfer function matrix, with $(I-P(q))$ invertible, $P$ being hollow, meaning that $[P(q)]_{kk} = 0$ for all $k$, and $Q$ being composed of a (sub)set of columns of the $m\times m$ identity matrix. Furthermore, and without loss of generality, it is assumed that $\Phi_x(\omega)$ is diagonal and $\Phi_x(\omega) \succ 0$ for all $\omega$. An example of a network graph that generates an input vector $u$ is shown in Figure \ref{fig1}, in the particular situation of a target module $G_{j1}$.

Note that in this representation $x$ can be either a user-controlled measured signal or an unmeasured disturbance process.\\
The reason for writing the input signal in this specific form, is that it allows us to specify dependencies among input signals in a graph-based format, which is particularly attractive when generalizing our framework from an open-loop situation to a dynamic network situation.

\begin{definition}
Consider a network graph $\mG$ with vertex set $\mV := \{u_k\}_{k=1,\cdots m} \cup \{x_k\}_{k=1,\cdots n_x}$ and directed edge/link set $\mE \subseteq \mV \times \mV$:
\begin{itemize}
\item $(u_k,u_{\ell}) \in \mE$ if and only if $[P]_{\ell k} \neq 0$;
\item $(x_k,u_{\ell}) \in \mE$ if and only if $[Q]_{\ell k} \neq 0$.
\end{itemize}
Vertex $u_{\ell}$ is an out-neighbour of $u_{k}$ if $(u_k,u_{\ell})\in \mE$. \hfill $\Box$
\end{definition}

\begin{definition}
A vertex set $\mathcal{V}_{\mD^c}$ is called a disconnecting set from $\mV_1$ to $\mV_2$, if upon removal of $\mathcal{V}_{\mD^c}$ from the graph, there is no directed path from $\mV_1$ to $\mV_2$. It is a minimum disconnecting set if it has the minimum cardinality among all such disconnecting sets. \hfill $\Box$
\end{definition}

Note that $\mV_{\mD^c}$ may contain elements of $\mV_1$ and/or $\mV_2$.

We recall that $\kappa(t) := \begin{bmatrix} u(t) \\ e(t) \end{bmatrix}$ and that $\kappa^{[j]}(t)$ is the selection of components in $\kappa(t)$ that correspond to parametrized entries in the columns of $M_j(q,\theta)$. In line with this decomposition of $\kappa(t)$, we write $\kappa^{[j]}$ as $\kappa^{[j]} = \begin{bmatrix} \kappa_u^{[j]}  \\ \kappa_e^{[j]}\end{bmatrix}$. This implies that the signals in $\kappa_u^{[j]}$ are the inputs to parametrized modules in the MISO model with output $y_j$.
Note that the input signals to modules that are known a priori are not present in $\kappa^{[j]}$.

As a next step we are going to decompose the input signals in $\kappa_u^{[j]}$ in three different components. In doing so we adopt the notation that for an $m$-dimensional node vector $u$, and $\mD^c \subset \{1,\cdots m\}$ an index set, $u_{\smD^c}$ is a vector of those components of $u$ that are defined by the index set $\mD^c$. We will use this general notation throughout the remaining part of the paper.
\begin{definition}
\label{def:disc}
Let $\{u_{\smD^c}\}$ be defined as a disconnecting set from $\{u_i\}$ to the other components in $\{\kappa_u^{[j]}\}$, under the constraints that $i \notin \mD^c$.\footnote{$u_{\smD^c}$ will obviously be dependent on $i$ and $j$. In the remaining text we will focus on one fixed choice of $i$ and $j$, avoiding the use of more complex notation.} \hfill $\Box$
\end{definition}
The interpretation of this disconnecting set is that it disconnects the input $u_i$ of the target module from all other inputs to parametrized modules. To illustrate this, consider the situation as depicted in Figure \ref{fig1}. The disconnecting set $\{u_2\}$ disconnects $\{u_1\}$ from $\{u_3\}$. As a result, the dependency of $u_3$ on $u_1$ is fully covered by the signal $u_2$. This phenomenon is closely related to the concept of blocked parallel paths from $u_1$ to $y_j$, i.e. parallel to the target module $G_{j1}$ \citep{Dankers&etal_TAC:16}, i.e. $u_2$ blocks all connections from $u_1$ to $y_j$ that are not passing through $G_{j1}$.

As we are focusing on estimating the entry $G_{ji}(q)$ of our MIMO system, we focus on its input $u_i$ and determine a decomposition of the vector $\kappa_u^{[j]}$ as follows:
\beq
\label{eq11}
   \kappa_u^{[j]} = \begin{bmatrix} u_i \\ u_{\smD^c} \\ u_{\smE} \end{bmatrix}
\eeq
where $u_{\smE}$ is a remainder set, after defining $u_{\smD^c}$. In the example of Figure \ref{fig1}, $u_i = u_1$, $u_{\smD^c} = u_2$ and $u_{\smE} = \begin{bmatrix} u_3 \\ u_4 \end{bmatrix}$.

The full input vector $u$ is then composed of the elements in $\kappa_u^{[j]}$ complemented with the input signals $u_{\smF}$ related to a priori known terms in $G(q,\theta)_{j*}$.
In the sequel the following set of external signals will appear to be instrumental.
\begin{definition}
\label{def5}
Let $x_{\smE^*}$ be defined as those components in vector signal $x$, that in the graph of the input signal $u$, have a path to $\red{\{}u_{\smE}\red{\}}$ without passing through $\red{\{}u_{\smD^c}\red{\}}$. \hfill $\Box$
\end{definition}
On the basis of expression (\ref{equ}) for $u$, we can then remove (immerse) the signals $u_{\smF}$ from the equations and derive an expression for $u_{\smE}$ as follows.
\begin{proposition}
\label{prop3}
Consider the expression (\ref{equ}) for $u$ and consider the decomposition of $u$ as specified in (\ref{eq11}). Then there exist rational transfer function matrices $T_s(q)$ and $R_s(q)$ such that $u_{\smE}$ can be written as:
\beq
\label{eqrs}
  u_{\smE}(t) = T_s(q) u_{\smD^c}(t) + R_s(q)x_{\smE^*}(t).
\eeq
\end{proposition}

{\bf Proof} See Appendix.

Note that for external signals that pass through $\red{\{}u_{\smD^c}\red{\}}$ before reaching $\red{\{}u_{\smE}\red{\}}$, their contribution to $u_{\smE}$ is covered by the transfer function $T_s(q)$ in (\ref{eqrs}).

We can now formulate the following result, which is a further sharpening of the result of Proposition \ref{prop2}.

\begin{theorem}
\label{theo1}
Consider the situation of Proposition \ref{prop2}, where the input signals are generated according to (\ref{equ}), the input signal decomposition (\ref{eq11}), and $x_{\smE^*}$ according to Definition \ref{def5}.
\begin{itemize}
\item Let the columns of $M_{j}(q,\theta)$ corresponding to the input signals $u_{i\cup\smD^c}$ be parametrized independently from the columns corresponding to the input signals $u_{\smE}$.
\item Let \red{$\subsuper{u}{i\cup\smD^c}{\perp x_{\smE^*}}$} be the projection\footnote{Projection is considered with respect to the inner product defined by $<\! x_m,y_n\! > := \Eb [x_m(t)y_n(t)]$, while the projection of a vector signal $x$ onto a vector signal $y$ is defined as the projection of every component of vector $x$ onto all components of vector $y$.}
    of signals $u_{i\cup\smD^c}$ onto the orthogonal complement of $x_{\smE^*}$.
\end{itemize}
Then $G_{ji}(q,\theta) = G_{ji}^0(q)$ is unique in the minimum of the cost function in (\ref{eq:ts}) if
\beq
\label{eqphic}
\Phi_{\mu^{[j]}}(\omega) \succ 0\ \mbox{for almost all}\ \omega
\eeq
with
\beq
\label{eqmuj}
\mu^{[j]} := \begin{bmatrix} \red{\subsuper{u}{i\cup\smD^c}{\perp x_{\smE^*}}} \\ \kappa_e^{[j]} \end{bmatrix}.
\eeq
\end{theorem}

{\bf Proof} See Appendix. \hfill $\Box$

The interpretation of this result is as follows:
\begin{enumerate}
\item[a.] It is not the full input vector $\kappa_u^{[j]}$ of inputs that have a parametrized map to the output, that requires excitation. Only a subset of input signals, present in $u_{i\cup\mD^c}$ need excitation.
\item[b.] For exciting this subset of inputs, all external signals $x$ can be used, except for the $x$ signals that are present in $x_{\smE^*}$. For excitation of $u_{i\cup\mD^c}$ we can use those signals $x$ that in the graph of the input network reach $\red{\{}u_{i\cup\mD^c}\red{\}}$ without passing through $\red{\{}u_{\smE}\red{\}}$.
\item[c.] The white noise signals in $\kappa_e^{[j]}$ are trivially sufficiently exciting.
\end{enumerate}

\begin{remark}
Since the white noise signals $e$ in $\kappa_e^{[j]}$ are considered to be independent of $x$, an equivalent statement for (\ref{eqphic}) would be $\Phi_{\red{\subsuper{u}{i\cup\smD^c}{\perp x_{\smE^*}}}}(\omega) \succ 0$. However, for generalization of these results to the dynamic network case, we prefer the current expressions (\ref{eqphic})-(\ref{eqmuj}).
\end{remark}

\begin{remark}
In this section, the node set $\{u_{\smD^c}\}$ is defined as any disconnecting set from $u_i$ to $\{\kappa_u^{[j]}\}\backslash \{u_i\}$. It may seem attractive to select it as a {\it minimum} disconnecting set to reduce the number of required excitation signals, but since $\mD^c$ also influences the set $x_{\smT^*}$ such a choice does not always lead to the smallest number of excitation signals to be added.
\end{remark}

Interpretation (a.) above confirms the earlier insights of \cite{Dankers&etal_TAC:16}, that for consistently estimating a target module $G_{ji}$ it would be sufficient to measure input $u_i$ and those inputs that block all parallel paths from $u_i$ to $y_j$ This has been formalized in the Parallel Path and Loop condition \citep{Dankers&etal_TAC:16}, or alternatively phrased as a separating/disconnecting set condition \citep{Dankers&etal_TAC:16,Shi&etal_Autom:22}. \\
The current results show that we do not need to exclude the remaining inputs from the predictor model, as was done in \cite{Dankers&etal_TAC:16}. We can include the inputs $u_{\smE}$ in our predictor model, but the concerned input signals do not require additional excitation. Their effect on output $y_j$ is modeled, but for this we do not require a consistent estimate of the concerned transfer functions that map the input signals to $y_j$.
In this setting the current result improves on the results of \cite{VandenHof&Ramaswamy_CDC:20,VandenHof&etal_IFAC:23} where persistence of excitation of all inputs was chosen in order to guarantee data-informativity.

\section{Example}
\label{sec:exam}
We illustrate the result of Theorem \ref{theo1} through the four-input one-output example as indicated in Figure \ref{fig1}. In the considered situation, $G_{j1}$ is the target module to be estimated, while the predictor model is given by:
\beq
   \varepsilon(t,\theta) = H(q,\theta)^{-1}[y_j(t) - \sum_{k=1}^4 G_{jk}(q,\theta)u_k(t)].
\eeq
First we consider the situation that all four modules $G_{jk}(q,\theta)$, $k=1,\cdots 4$ are parametrized.
A minimum disconnecting set $\{u_{\smD^c}\}$ from input $u_i=u_1$ to all other inputs of parametrized modules is $\{u_2\}$. In an alternative interpretation it is clear that $u_2$ blocks all parallel paths (i.e. parallel to the target module $G_{j1}$) from $u_i$ to output $y_j$.\\
The signal $u_{i\cup\smD^c}$ is composed of $u_i = u_1$, $u_{\smD^c} = u_2$, while $u_{\smE} = \red{[u_3\ u_4]}^T$. Possible excitation or noise signals $x_3$ and $x_4$ will serve as $x_{\smE^*}$. Since $u$ and $e$ are considered to be independent, and $e$ is white noise, the condition (\ref{eqphic}) can be replaced by a positive definite spectrum of \red{$\subsuper{u}{\{1\cup2\}}{\perp \{x_3,x_4\}}$}.\\
For evaluating this spectrum we can write $u_1$ and $u_2$ as function of the external signals affecting the system:
\beq
  u_1  =  \sum_{k=1}^4 F_{1k}(q) x_k(t);\ \ \ \
  u_2 =  \sum_{k=1}^4 F_{2k}(q) x_k(t).
\eeq
The vector signal \red{$\subsuper{u}{\{1\cup2\}}{\perp \{x_3,x_4\}}$} can then be written as
\[
\red{\subsuper{u}{\{1\cup2\}}{\perp \{x_3,x_4\}}} = \underbrace{\begin{bmatrix} F_{11}(q) & F_{12}(q) \\ F_{21}(q) & F_{22}(q) \end{bmatrix}}_{F(q)} \begin{bmatrix} x_1 \\ x_2\end{bmatrix}
\]
A positive definite spectrum of \red{$\subsuper{u}{\{1\cup2\}}{\perp \{x_3,x_4\}}$} is then obtained for almost all $\omega$ if the matrix $F(z)$ has full row rank over the field of rational functions, and the spectrum of $[x_1\ \ x_2]^T$ is positive definite
\cite{VandenHof&Ramaswamy_CDC:20}.

This shows that excitation of inputs $u_1$ and $u_2$ is sufficient for data-informativity, while excitation of $u_3$ and $u_4$ is not necessary.\\
Moreover the signals $x_3$ and $x_4$ cannot be used to excite $(u_1,u_2)$ since they pass through $\red{\{}u_{\smE}\red{\}}$ before reaching $\red{\{}u_2\red{\}} (= \red{\{}u_{\smD^c}\red{\}})$, and therefore they belong to $x_{\smE^*}$ . To understand why $x_3$ and $x_4$ cannot be used for excitation of $(u_1,u_2)$ we consider the prediction error
\[
\varepsilon(t,\theta) = H(q,\theta)^{-1} \left[ \sum_{k=1}^4 \Delta G_{jk}(q,\theta)u_k(t) + H^{0}(q)e(t)\right].
\]
Writing $u_k$ as a function of the variables $x$, according to the situation of Figure \ref{fig1}, this can be rewritten as
\begin{equation}
\varepsilon(t,\theta) = H(q,\theta)^{-1} \left[ \sum_{k=1}^4 T_{k}(q,\theta)x_k(t) + H^{0}(q)e(t)\right] \nonumber
\end{equation}
with
\begin{eqnarray*}
T_1(q,\theta) & = & \Delta G_{j1}+\Delta G_{j2}\frac{P_{21}}{1\!-\!P_{32}P_{23}}\! +\! \Delta G_{j3}\frac{P_{32}P_{21}}{1\!-\!P_{32}P_{23}} \\
T_2(q,\theta) & = & \Delta G_{j2}\frac{1}{1-P_{32}P_{23}} + \Delta G_{j3}\frac{P_{32}}{1-P_{32}P_{23}} \\
T_3(q,\theta) & = & \Delta G_{j2}\frac{P_{23}}{1-P_{32}P_{23}} + \Delta G_{j3}\frac{1}{1-P_{32}P_{23}} \\
T_4(q,\theta) & = & \Delta G_{j2}\frac{P_{23}P_{34}}{1\!-\!P_{32}P_{23}}\! +\! \Delta G_{j3}\frac{P_{34}}{1\!-\!P_{32}P_{23}}\! +\! \Delta G_{j4}
\end{eqnarray*}
where $\Delta G_{jk} := G_{jk}^{0}(q)-G_{jk}(q,\theta)$, for $k = 1,\cdots 4$.  The minimum of the cost function $\Eb \varepsilon^T(t,\theta)\varepsilon(t,\theta)$ is achieved if the $T_k$-dependent terms are zero. Presence of $x_k$ implies that $T_k(q,\theta) = 0$ in the minimum of the cost function. If $x_1$ and $x_2$ are present and persistently exciting, then $T_1$ and $T_2$ are forced to be zero, which implies that $\Delta G_{j1} = 0$, and thus a consistent estimate of $G^{0}_{j_1}$ is guaranteed. This follows from the observation that $T_1(q,\theta) = \Delta G_{j1}+ P_{21}T_2(q,\theta)$. If we would replace $x_2$ by presence of $x_3$, a similar result does not exist. Introducing $x_4$ shows that $T_4(q,\theta)$ gets involved in the expression and a new unknown term $\Delta G_{j4}$ is introduced also. This explains that forcing $T_1$ and $T_4$ to be zero is not sufficient for arriving at $\Delta G_{j1} = 0$, or in other words: exciting $x_1$ and $x_4$ does not lead to a consistent estimate of the target module.

In the case that module $G_{j2}(q)$ is not parametrized, but known a priori, there are two options for choosing (minimum) disconnecting sets:
\begin{itemize}
\item When selecting  $u_{\smD^c} = u_2$, it follows that $u_{\smE} = \red{[u_3\ u_4]}^T$, and as a result $x_3$ and $x_4$ cannot be used for excitation of $(u_1,u_2)$, and therefore we need excitations $x_1$ and $x_2$.
\item When selecting $u_{\smD^c} = u_3$, it follows that $u_{\smE} = u_4$, and as a result $x_4$ cannot be used for excitation of $(u_1,u_3)$. This implies that external signals $(x_1,x_2,x_3)$ can be used for realizing a positive definite spectrum of \red{${u_{\{ 1 \cup 3\}}}\!^{\perp x_{\smE^*}}$}. This is guaranteed by having an excitation $x_1$ plus either $x_2$ or $x_3$.
\end{itemize}
The result can be verified by setting $\Delta G_{j2} = 0$ in the above expressions for $T_k(q,\theta)$. Then $\Delta G_{j3} = 0$ will follow from the presence of either $x_2$ or $x_3$, while $\Delta G_{j1}$ then follows through the presence of $x_1$.

Note that in case multiple disconnecting sets are possible, each one of them provides a set of sufficient conditions for data-informativity. Not satisfying the conditions for one particular choice of disconnecting set, does not necessarily lead  to the conclusion that the data-informativity conditions are not satisfied.

In the case that $G_{j4}$ is not parametrized and known a priori, the (minimum) disconnecting set from $u_1$ to $\{u_2,u_3\}$ is given by $u_2$. Then $u_{\smE} = u_3$, while the external signals $x_{\smE^*}$ that affect $u_{\smE}$ are given by $x_{\smE^*} = \red{[x_3\ x_4]}^T$. This implies that (only) signals $(x_1,x_2)$ are available for realizing a positive definite spectrum of \red{${u_{\{1 \cup 2\}}}\!^{\perp x_{\smE^*}}$}. Verification of this through the expressions of $T_k(q,\theta)$, shows that with $\Delta G_{j4}=0$, the linear combination of the two terms in $T_3$ and $T_4$ are different from the linear combinations of $\Delta G_{j2}$ and $\Delta G_{j3}$ in the expressions for $T_1$ and $T_2$. As a result the original result with all modules parametrized follows as the solution.

\section{Path-based interpretation}
\label{sec:path}
As shown in earlier work \cite{VandenHof&Ramaswamy_CDC:20}, a spectrum condition for data-informativity like (\ref{eqphic}), can be guaranteed to hold generically, if particular path-based conditions on the network graph are satisfied. As was shown to hold in a related identifiability studies in \cite{Hendrickx&Gevers&Bazanella_TAC:19}, rank conditions on particular transfer functions in a dynamic network can generically be translated into conditions on vertex disjoint path connections between sets of nodes in the network graph.
Two directed paths are called vertex disjoint if they do not
share any vertex, including the starting and ending vertices.
They can be calculated on the basis of the network graph through a max-flow min-cut algorithm as the Ford-Fulkerson algorithm. For handling this result we copy the formulation from \cite{VandenHof&etal_IFAC:23}, replacing $\kappa$ by $\mu^{[j]}$ from (\ref{eqmuj}):
\begin{proposition}[\cite{VandenHof&etal_IFAC:23}]
\label{propg}
The spectrum condition $\Phi_{\mu^{[j]}} \succ 0$ for almost all $\omega$, holds generically\footnote{In line with its use in \cite{Hendrickx&Gevers&Bazanella_TAC:19}, this concerns a property that holds for all possible dynamic modules in the network, except possibly for a set of measure $0$.} if in the graph of the network there exist $dim(\mu^{[j]})$ vertex disjoint paths from the node sets $\{x,e\}$ to $\{\mu^{[j]}\}$, and the signals in $x$ are mutually independent and persistently exciting.
\end{proposition}
Actually, it should be sufficient to require only those $x$ signals to be persistently exciting that serve as starting nodes in the vertex disjoint paths.
When specifying the above result for the particular situation treated in Section \ref{sec:di4ss}, we obtain the following.
\begin{corollary}
The graph-based condition in Proposition \ref{propg} is satisfied if 
there exist $dim(u_{\smD^c})+1$ vertex disjoint paths between the node sets $\{x \backslash x_{\smE^*}\}$ and $\{u_{i\cup\smD^c}\}$.
\end{corollary}
{\bf Proof.} Since $u$ and $e$ are independent, the spectrum condition $\Phi_{\mu^{[j]}}(\omega) \succ 0$ can be recast into the expression $\Phi_{\red{\subsuper{u}{i\cup\smD^c}{\perp x_{\smE^*}}}}(\omega) \succ 0$. Together with the fact that by assumption $\Phi_x(\omega) \succ 0$ for all $\omega$, this directly leads to the result. \hfill $\Box$

We can illustrate this result for the example in Section \ref{sec:exam}. In the situation where all $G$-modules are parametrized it was found that $\{u_{\smD^c}\} = \{ u_2\}$, $\{u_{i\cup\smD^c}\} = \{u_1,u_2\}$ and $\{x \backslash x_{\smE^*}\} = \{x_1,x_2\}$. Therefore, two vertex disjoint paths between $\{x_1,x_2\}$ and $\{u_1, u_2\}$, are sufficient to guarantee (generic) data-informativity. This condition is satisfied if both $x_1$ and $x_2$ are present.

\section{Single module identification in dynamic networks}
\label{sec:dynnet}
\subsection{Single module data-informativity results}
\label{eq81}
In this section we are going to make the step from the isolated open-loop situation as considered before, to the situation where the considered MIMO/MISO system can be part of a dynamic network, including possible feedback loops. Dynamic networks are interconnections of dynamic systems and can be represented in different forms, one of them being the so-called module representation \citep{VandenHof&etal_Autom:13}, defined by
\begin{equation}
\label{eq:orig}
  w(t) = G^0(q) w(t) + H^0(q)e(t) + u(t)
\end{equation}
with $w$ an $L$-dimensional vector, $G^0(q)$ a hollow rational transfer function matrix, i.e. with zeros on the diagonal entries,
$e$ an $L$-dimensional vector of white noise processes, $H^0(q)$ the rational disturbance model, and
$u(t) = R^0 \cdot r(t)$, accounting for the effect of measured external excitation signals $r$ on the network, with $R^0$ a binary $L \times K$ matrix and $r$ a $K$-dimensional vector of excitation signals. A non-zero element $G_{k\ell}^0(q)$ in $G^0(q)$ is referred to as a {\it module}.\\
It is further assumed that the network is stable and well-posed \citep{VandenHof&etal_Autom:13}. In this paper it is also assumed, for simplicity, that all elements in $G^0(q)$ are strictly proper.

When targeting on the identification of a single module $G_{ji}^0$ in this network, several approaches are available for arriving at a consistent module estimate. An indirect identification method is suitable when there is a sufficient number of $u$ signals present in the network, see e.g
\cite{VandenHof&etal_Autom:13,Dankers&etal_TAC:16,Gevers&etal_sysid:18}. When noise signals $e$ are required to provide sufficient excitation in the network for consistent estimation of the target module, a direct method is the prime approach, see e.g. \cite{VandenHof&etal_Autom:13,Dankers&etal_TAC:16,Ramaswamy&VandenHof_TAC:21}. In this paper we will focus on the direct method.

In this method a predictor model is constructed composed of a set of predictor inputs $w_{\smD}$ and a set of predicted outputs $w_{\smY}$, on the basis of which the following predictor model equation can be formulated:
\beq
\label{eq:pm}
 w_{\smY}(t) = \bar G^0(q) w_{\smD}(t) + \bar H^0(q) \xi_{\smY}(t) + \bar T^0(q) u(t).
\eeq
This actually concerns the description of part of the network (\ref{eq:orig}), where the equations for the nodes outside of  $w_{\smY}$ have been decoupled and removed,
while the network nodes that are not present in $w_{\mY\cup\mD}$ have been removed (immersed) from the description. The noise contribution $\bar H^0(q) \xi_{\smY}(t)$, represents the (correlated) noise that affects the output $w_{\smY}$, with $\bar H^0(q)$ being monic, stable and stably invertible, and $\xi_{\smY}$ a vector white noise process. In general, the noise model in (\ref{eq:pm}) is the result of a spectral factorization, after manipulating the network equation (\ref{eq:orig}); that is why the noise term $\xi_{\smY}$ is typically different from $e_{\smY}$. The matrix $\bar T^0(q)$ represents the effect of external excitation signals on $w_{\smY}$. Note that not all excitation signals that are present in the network will appear in the term $\bar T^0(q) u(t)$. Excitation signals that enter the network on nodes $w_{\smD}$ will generally contribute to $w_{\smY}$ through their respective input term $w_{\smD}$. The matrix $\bar T^0(q)$ can have both dynamic terms and fixed terms, the latter situation, e.g., occurring when an excitation signal directly enters the network on a node in $w_{\smY}$.

The single module identification problem is then formulated as the problem of identifying a single module $\bar G_{ji}^0(q)$, defined as the mapping from $w_i$ to $w_j$ in (\ref{eq:pm}), on the basis of data $\{w_{\smD\cup\smY}(t),u(t)\}_{t=0,\cdots N-1}$, under the condition that $\bar G_{ji}^0(q) = G_{ji}^0(q)$, i.e. the original module in the network.
Estimation is done on the basis of the one-step ahead prediction error
\beq
\label{eq:pe}
\varepsilon(t,\theta) = \bar H(q,\theta)^{-1} \left[ w_{\smY}(t) - \bar G(q,\theta)w_{\smD}(t) - \bar T(q,\theta)u(t)\right]
\eeq
and a quadratic type of cost function as in (\ref{eq:ts}).

In order to arrive at a consistent estimate (\cite{Ljung:99}) of the target module, a set of  conditions has been formulated in previous works:
\begin{enumerate}
\item[a.] In order to guarantee module invariance, i.e. $\bar G_{ji}^0(q) = G_{ji}^0(q)$, a parallel path and loop (PPL) condition\footnote{The PPL condition \cite{Dankers&etal_TAC:16} requires that each path in the network from $w_i$ to $w_j$ excluding the path through $G_{ji}$, and every loop around $w_j$, pass through a measured node that is included in the predictor input.} is assumed to be satisfied, which influences the selection of predictor inputs $w_{\smD}$, see \cite{Dankers&etal_TAC:16}.
\item[b.] In order to cope with {\it confounding variables}\footnote{A confounding variable is an unmeasured variable that has paths to both the
input and output of an estimation problem \cite{Pearl:2000}.} multivariable noise models can be used to model the correlated disturbances, leading to a selection of an appropriate set of node signals in $w_{\smY}$ and $w_{\smD}$, see \cite{Ramaswamy&VandenHof_TAC:21}.
\item[c.] Conditions on data-informativity, see \cite{VandenHof&etal_IFAC:23}.
\item[d.] Technical conditions on the presence of delays in the network, see \cite{Ramaswamy&VandenHof_TAC:21}.
\item[e.] A system-in-the-model set condition, implying that the chosen model set is sufficiently flexible to include the true system dynamics.
\end{enumerate}

In the research so far, the mentioned data-informativity conditions have been formulated for consistent estimation of the full ``subnetwork'' $\bar G^0(q)$. However, in view of the results presented in Section \ref{sec:di4ss}, these conditions can now be further relaxed, by focusing on consistent estimation of $\bar G_{j\star}^0(q)$ only, i.e. the $j$-th row of $\bar G^0(q)$, corresponding to the output of the target module, and actually on element $\bar G_{ji}^0(q) = G_{ji}^0(q)$ only, while taking account of the structurally zero and possible fixed (non-parametrized) entries in this row.

For the formulation of the relaxed conditions, we denote
\beq
\label{eq:mj}
M(q,\theta) = \begin{bmatrix} \bar G(q,\theta) & \bar H(q,\theta) & \bar T(q,\theta) \end{bmatrix}
\eeq
and $\Delta M(q,\theta) =
\begin{bmatrix} \Delta \bar G(q,\theta) & \Delta \bar H(q,\theta) & \Delta \bar T(q,\theta) \end{bmatrix}$, with $\Delta \bar G(q,\theta) := \bar G^0(q) - \bar G(q,\theta)$ etcetera, where the resulting predict\red{ion} error can be written as
\beq
  \varepsilon(t,\theta) = \bar H(q,\theta)^{-1}\Delta M(q,\theta) \kappa(t)
\eeq
with
\beq
\kappa = \begin{bmatrix} w_{\smD} \\ \xi_{\smY} \\ u \end{bmatrix}.
\eeq
In accordance with the reasoning in Section \ref{sec:di4ss} we focus on output node $w_j$ and therefore on the $j$-th row $M_j(q,\theta)$ of $M(q,\theta)$, evaluating under which conditions the power of the signal $\Delta M_j(q,\theta)\kappa^{[j]}$ is equal to zero, where
\beq
\label{eqkj}
  \kappa^{[j]} = \begin{bmatrix} w_{\smD}^{[j]}\\ \xi_{\smY}^{[j]} \\ u^{[j]} \end{bmatrix}
\eeq
being those terms in $\kappa$ that correspond with a {\it parametrized} term in the corresponding column of the row $M_j(q,\theta)$.

\begin{definition}
\label{def:disc2}
Let $\{w_{\smD^c}\}$ be defined as a disconnecting set from $w_i$ to the other components in $\{w_{\smD}^{[j]}\}$, under the constraint that $i \notin \mD^c$.
\end{definition}

Note that Definition \ref{def:disc2} is similar to Definition \ref{def:disc}, but the consequence is slightly different, in the sense that, because of the network structure, there can be nodes in $w_{\smD^c}$ now that are not necessarily in-neighbors of $w_j$. Similar to the situation of Section \ref{sec:di4ss}, we define
\begin{itemize}
\item $w_{\smE}$ as the node vector defined by $w_{\smE} := w_{\smD\red{\backslash\{w_i\cup w_{\smD^c} \}}}^{[j]}$, i.e. the set of input nodes to parametrized modules, excluding $w_i$, that are not in the disconnecting set.
\item $w_{\smF}^{[j]}$ as the node vector with nodes that are inputs to a non-parametrized (known) module in the $j$-th row of the predictor model.
\end{itemize}

We can now formulate a companion result for Proposition \ref{prop3} for the network situation.

\begin{definition}
\label{def6}
Let $x_{\smE^*}$ be defined as those components in vector signal $x = vec(u,e)$, that in the graph of the network, have a path to $\red{\{}w_{\smE}\red{\}}$ without passing through $\red{\{}w_{\smD^c}\red{\}}$.
\end{definition}

\begin{proposition}
\label{prop7}
Consider the network equation (\ref{eq:orig}) for $w$ and consider the node vectors $w_i, w_{\smE}, w_{\smD^c}$ and $x_{\smE^*}$ as defined above. Then there exist rational transfer function matrices $T_s(q)$ and $R_s(q)$ such that $w_{\smE}$ can be written as:
\beq
\label{eqrs2}
  w_{\smE}(t) = T_s(q) w_{\smD^c}(t) + R_s(q)x_{\smE^*}(t).
\eeq
\end{proposition}

{\bf Proof} The proof of this result is dual to the proof of Proposition \ref{prop3}, with all $u$-node signals replaced by $w$-node signals. \hfill $\Box$

This provides the right conditions for showing the data-informativity result for single module identification.
\begin{theorem}
\label{theox}
Consider the network (\ref{eq:orig}), predictor model (\ref{eq:pm})-(\ref{eq:pe}), the signal vector $\kappa^{[j]}$ (\ref{eqkj}) and the node vectors $w_i, w_{\smD^c}, w_{\smE}, x_{\smE^*}$ as defined above.
If
\begin{itemize}
\item row $M_j(q,\theta)$ is parametrized independently from the other rows in $M(q,\theta)$, and
\item the columns in $M_j(q,\theta)$ corresponding to the inputs $w_{i\cup\smD^c}$ are parametrized independently from the other columns in $M_j(q,\theta)$,
\end{itemize}
then $\bar G_{ji}(q,\theta) = \bar G_{ji}^0(q)$ is unique in the minimum of the quadratic cost function (\ref{eq:ts}) applied to (\ref{eq:pe}), if
\begin{enumerate}
\item[(a)] $\Phi_{\eta^{[j]}}(\omega) \succ 0$, for almost all $\omega$
\end{enumerate}
with
\beq
\label{eqeta}
 \eta^{[j]} = \begin{bmatrix} \red{\subsuper{w}{i\cup \smD^c}{\perp\chi}}   \\ \xi_{\smY}^{[j]} \end{bmatrix}
\eeq
and
\beq
\label{eqchi}
\chi = \begin{bmatrix} x_{\smE^*} \\ u^{[j]} \end{bmatrix}.
\eeq
\end{theorem}

{\bf Proof} See Appendix. \hfill $\Box$

One of the main differences with the situation of Section \ref{sec:di4ss} is that the two components in the vector $\eta^{[j]}$ are no longer independent. Because of possible feedback loops present in the network, the concerned signals can be correlated. An implication of the spectrum condition is that $w_{i\cup \smD^c}$ should be persistently excited, through the presence of external signals that exclude $\chi$ and $\xi_{\smY}^{[j]}$, while $\xi_{\smY}^{[j]}$ should be persistently exciting in itself. This latter condition is due to the fact that these signals are inputs to parametrized terms in the noise model, and therefore cannot be used for exciting the elements in $G$.

The external signals $\chi$ that are excluded from serving as excitation sources for estimating $G_{ji}^0(q)$, now also include all $u$-signals that are input to a parametrized module towards the target output $w_j$, i.e. the inputs that correspond to parametrized terms in $[\bar T(q,\theta)]_{j*}$, the $j$-th row of $\bar T(q,\theta)$ in (\ref{eq:pe}). Note however that, different from the term $\xi_{\smY}^{[j]}$, the signals in $u^{[j]}$ do not require excitation, as they are not part of the vector signal $\eta^{[j]}$.

Following the reasoning in Section \ref{sec:path} we can now also formulate the path-based conditions for generically satisfying the positive definite spectrum condition in Theorem \ref{theox}.

\begin{proposition}
\label{propgraph2}
Consider the situation of Theorem \ref{theox}, and consider the selections of external signals, defined by:
\begin{itemize}
\item $e^{\perp \smY}$: All $e$ signals in the network that do not have an unknown path\footnote{An ``unknown path'' is a path that contains a link with unknown dynamics that is parametrized in the predictor model.} to $w_{\smY}$, and
\item $u^{\perp j}$: All $u$-signals in the network that do not have a parametrized link to $w_j$ in the predictor model.
\end{itemize}
Then the data informativity condition (a) in Theorem \ref{theox} is generically satisfied if there exist $dim(\mD^c)+1$ vertex disjoint paths from $\{e^{\perp\smY},u^{\perp j}\}\backslash\{x_{\smT^*}\}$ to
$\{w_{i\cup\smD^c}\}$, and the signals in $\{e^{\perp\smY},u^{\perp j}\}$ are mutually independent and persistently exciting.
\end{proposition}

{\bf Proof} See Appendix. \hfill $\Box$

The result shows that for appropriately exciting $w_{i\cup\smD^c}$, we can use all external signals that in the predictor model do not have a parametrized link to the output. External signals that do have such a parametrized link are required for exciting the concerned entries in $\bar H(q,\theta)$ and $\bar T(q,\theta)$ respectively, and therefore are not available for exciting the node inputs for estimating $G_{ji}^0(q)$. From the remaining set of external signals, only those external signals can be used that reach the predictor input signals through the set of nodes $\{w_{i\cup\smD^c}\}$, without passing through $\{w_{\smE}\}$.

\begin{figure}[h]
\centerline{\includegraphics[width=0.9\columnwidth]{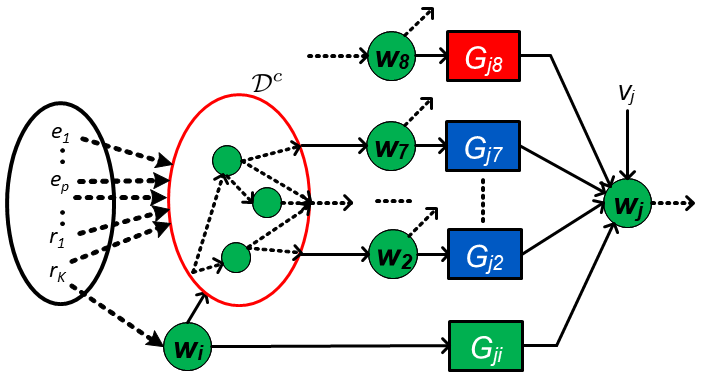}}
	\caption{Network situation with target module $G_{ji}$. Module $G_{j8}$ is known and therefore not parametrized. The disconnecting set $\{w_{\smD^c}\}$ is disconnecting from node $w_i$ to the remaining inputs to parametrized modules. $w_{\smD^c}$ and $w_i$ need sufficient excitation from external signals, either $e$- or $r-$ signals, while external signals that have a parametrized link to the output $w_j$ are excluded. Note that $u = R^0 r$ with $R^0$ a binary matrix.}
	\label{figx}
\end{figure}

An illustration of this situation is provided in Figure \ref{figx}.

\begin{remark}
The selection $e^{\perp\smY}$ as formulated in Proposition \ref{propgraph2}, can in some specific cases of predictor models, be slightly enlarged, in particular when the number of $e$ nodes with an unknown path to $e_{\smY}$ exceeds $dim(w_{\smY})$. This holds for predictor models that allow for predictor inputs that have a confounding variable with an output signal, a so-called nonempty set $\red{\{}w_{\smB}\red{\}}$ \cite{Ramaswamy&VandenHof_TAC:21}. A further specification of this will be presented elsewhere.
\end{remark}

\subsection{Discussion}

The new single module data-informativity results presented in Theorem \ref{theox} and Proposition \ref{propgraph2}, are relaxed versions of the data-informativity conditions presented in \cite{VandenHof&etal_IFAC:23}, where the target was to consistently estimate the full (unstructured) predictor model. It has been shown in \cite{VandenHof&etal_IFAC:23} that for satisfying those data-informativity conditions it may be necessary to adjust the selection of predictor inputs and outputs. In other words, for a particular selection of predictor inputs and outputs, it is not guaranteed that the data-informativity can be met by adding a sufficient number of external excitation signals. That is different in the current results, where the data-informativity conditions can always be satisfied by adding a sufficient number of external excitation signals at appropriate locations.

\begin{corollary}
\label{cor10}
Consider the situation of Theorem \ref{theox}. Then for any predictor model that satisfies the structural conditions for module invariance and confounding variables handling\footnote{See conditions a. and b. in Section \ref{eq81}.}, there exists a disconnecting set $\red{\{}w_{\smD^c}\red{\}}$ such that the data-informativity condition for consistent estimation of the target module $\bar G_{ji}^0(q)$ can be satisfied, by adding a sufficient number of external excitation signals to the predictor inputs $w_{i\cup\smD^c}$.
\end{corollary}

When estimating a single module $G_{ji}^0(q)$ in a dynamic network, the user has the option now to require two different sets of conditions for data-informativity, either the single module conditions of the current paper, or the full predictor model conditions of \cite{VandenHof&etal_IFAC:23}. While the single module conditions are more relaxed, this situation also comes at a cost. When validating an estimated module, as part of a consistently estimated predictor model, model validation can be applied to the full predictor model, for validating all of the present modules. This can e.g. be done by considering a validation data set. However when only the target module is estimated consistently, validation of this module using a validation data set, would require that in this validation data set, the node signals that are input to parametrized modules in the predictor model, are exactly the same as in the estimation data set.  This is caused by the fact that in the consistent single module estimation case, the estimated modules, different from the target module, will become dependent on the particular input node signals, present in the estimation data set.

So far we have been dealing with the data-informativity conditions for the direct method of estimating predictor models in a dynamic network, see \cite{Ramaswamy&VandenHof_TAC:21}. There are alternative methods, as e.g. the multi-step method of \cite{Fonken&etal_CDC:23}, where data-informativity conditions are also more relaxed.
Construction of predictor models for both the direct method and the multi-step method, including the different types of data-informativity conditions, has been implemented in the MATLAB app and toolbox SYSDYNET \cite{VandenHof&etal_SYSID:24}. 

One of the prime consequences of the results presented in this paper is in the relation with single module  identifiability. Identifiability conditions for single modules in a dynamic network have been studied in \cite{Weerts&etal_Autom:18_identifiability,Hendrickx&Gevers&Bazanella_TAC:19,Shi&etal_Autom:22,Shi&etal_TAC:23}. These conditions are typically not dependent on a particular identification algorithm. At the same time, identification algorithms have been studied that provide consistent target module estimates under specific conditions, among which data-informativity conditions. One would expect that if the single module identifiability conditions are satisfied, that there exists an estimation algorithm that can provide the consistent target module estimate. However, so far this step has only been achieved for a particular situation, namely when the single module identifiability conditions are satisfied by external excitation signals $r$ only. In that situation an indirect method can be used to estimate the target module consistently, see e.g. \cite{VandenHof&etal_Autom:13,Dankers&etal_TAC:16,Gevers&etal_sysid:18}. However when both $r$-excitation and external $e$-signals are required for satisfying the single module identifiability conditions, there was a gap with the related data-informativity conditions of the estimation algorithms.

In order to understand how the current results close this gap, we consider the identifiability result as formulated in Theorem 4 and Corollary 1 of \cite{Shi&etal_Autom:22}, adapted here to the situation of single module identifiability. We also consider a model set $\M$ of a full network that satisfies the basic assumptions as formulated in \cite{Shi&etal_Autom:22}, including the assumption that all modules in $G(q,\theta)$ are strictly proper.

\begin{theorem}[\cite{Shi&etal_Autom:22}]
\label{theo:shi}
Let $\mX_j$ be composed of those components in $x = vec(u,e)$ that in the model set $\M$ do not have a parametrized link to $w_j$. Then
$G_{ji}$ is generically identifiable in $\M$ from $(w,r)$ if
there exists a disconnecting set $\{w_{\smD^c}\}$ from $\{\mX_j \cup w_i\}$ to $\{w_{\smD}^{[j]}\}\backslash\{w_i\}$, subject to $i \notin \mD^c$, such that there exist $dim(w_{\smD^c})+1$ vertex disjoint paths from $\mX_j$ to $\{w_{i\cup\smD^c}\}$.
\end{theorem}

Now it appears that the conditions for generic single module data-informativity, as formulated in Proposition \ref{propgraph2} are extremely closely related to the generic single module identifiability conditions of Theorem \ref{theo:shi}. While in Proposition \ref{propgraph2} external signals in set $\red{\{}x_{\smE^*}\red{\}}$ are excluded from excitation, this similar situation is covered in Theorem \ref{theo:shi} by requiring that the disconnecting set $\red{\{}w_{\smD^c}\red{\}}$ should also be a disconnecting set from $\mX_j$ to $\{w_{\smD}^{[j]}\}\backslash \{w_i\}$. I.e. external signals that reach $\{w_{\smD}^{[j]}\}\backslash \{w_i\}$ without passing through $\{w_{\smD^c}\}$ are excluded. This set is exactly covered by $\red{\{}w_{\smT}\red{\}}$. \\
The single difference between the conditions in the two results, is in terms of the white noise signals $e$ that are available for excitation. While in Theorem \ref{theo:shi} only white noise signals with a parametrized link to $w_j$ are excluded, in Proposition \ref{propgraph2} these are the white noise signals with a parametrized path to $\red{\{}w_{\smY}\red{\}}$. In the situation of a single output (MISO) predictor model these sets are exactly the same. However in the multi-output case, there is apparently some conservatism in the direct identification method. Choosing a multiple output predictor model for identification of a single module is typically done for coping with confounding variables, see e.g. \cite{Ramaswamy&VandenHof_TAC:21}. Increasing the number of outputs apparently comes at the cost of more strict data-informativity conditions. This is in line with the data-informativity conditions for consistency of full predictor models, as presented in  \cite{Ramaswamy&VandenHof_TAC:21,VandenHof&etal_IFAC:23}. The multi-step method (\cite{Fonken&etal_CDC:23}) does not seem to suffer from this conservatism, due to an alternative handling of confounding variables.

\section{Examples}
\label{sec:ex2}
\subsection{Two-node example}

\begin{figure}[htb]
\centerline{\includegraphics[scale=0.7]{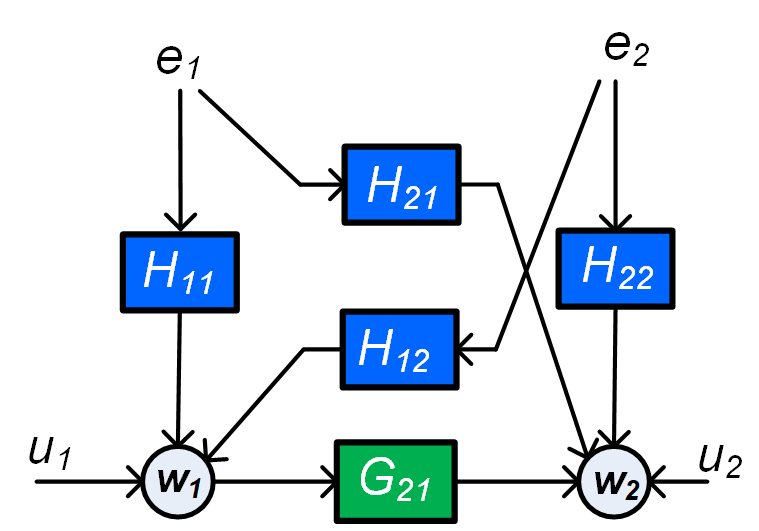}}
	\caption{Two node example with correlated disturbances and target module $G_{21}(q)$.}
	\label{fig2node}
\end{figure}

We consider the 2-node example as depicted in Figure \ref{fig2node} \citep{VandenHof&Ramaswamy_CDC:20} with target module $G_{21}$.

Because of the confounding variables that occur due to the correlated disturbances, we include node signal $w_1$ in the output, leading to a predictor model $w_1 \rightarrow (w_1,w_2)$. In the original reasoning of the local direct method, data-informativity is analyzed for the full SIMO model, leading to the situation that neither $u_1$ nor $u_2$ can be used for excitation of input $w_1$.
This is due to the fact that in modelling output $w_1$ in the predictor model, and since $w_2$ is not an input signal, both $u_1$ and $u_2$ appear in the predictor expression for $w_1$ with unknown dynamic terms, reflecting the sensitivity of the feedback loop between $w_1$ and $w_2$. The solution was then to move to a 2-input 2-output model $(w_1,w_2)\rightarrow (w_1,w_2)$ which then satisfies the data-informativity conditions when both $u_1$ and $u_2$ are present \cite{VandenHof&etal_IFAC:23}. In the analysis of the current paper, we consider the predictor model $w_1 \rightarrow (w_1,w_2)$, but for data-informativity only need to account for the mapping $w_1 \rightarrow w_2$. In this case $w_i = w_1$, $w_j = w_2$, $\mT = \emptyset$ and $\mD^c = \emptyset$. Since in the predictor model for predicting $w_2$, $u_1$ is not present, and $u_2$ appears with a constant term $1$, it follows that $u^{[j]} = \emptyset$, and therefore both $u_1$ and $u_2$ can be used to excite $w_1$. As a result, data-informativity is guaranteed generically if either $u_1$ or $u_2$ is present.

\subsection{Six-node example}
\label{sec:ex22}
\begin{figure}[htb]
\centerline{\includegraphics[scale=0.45]{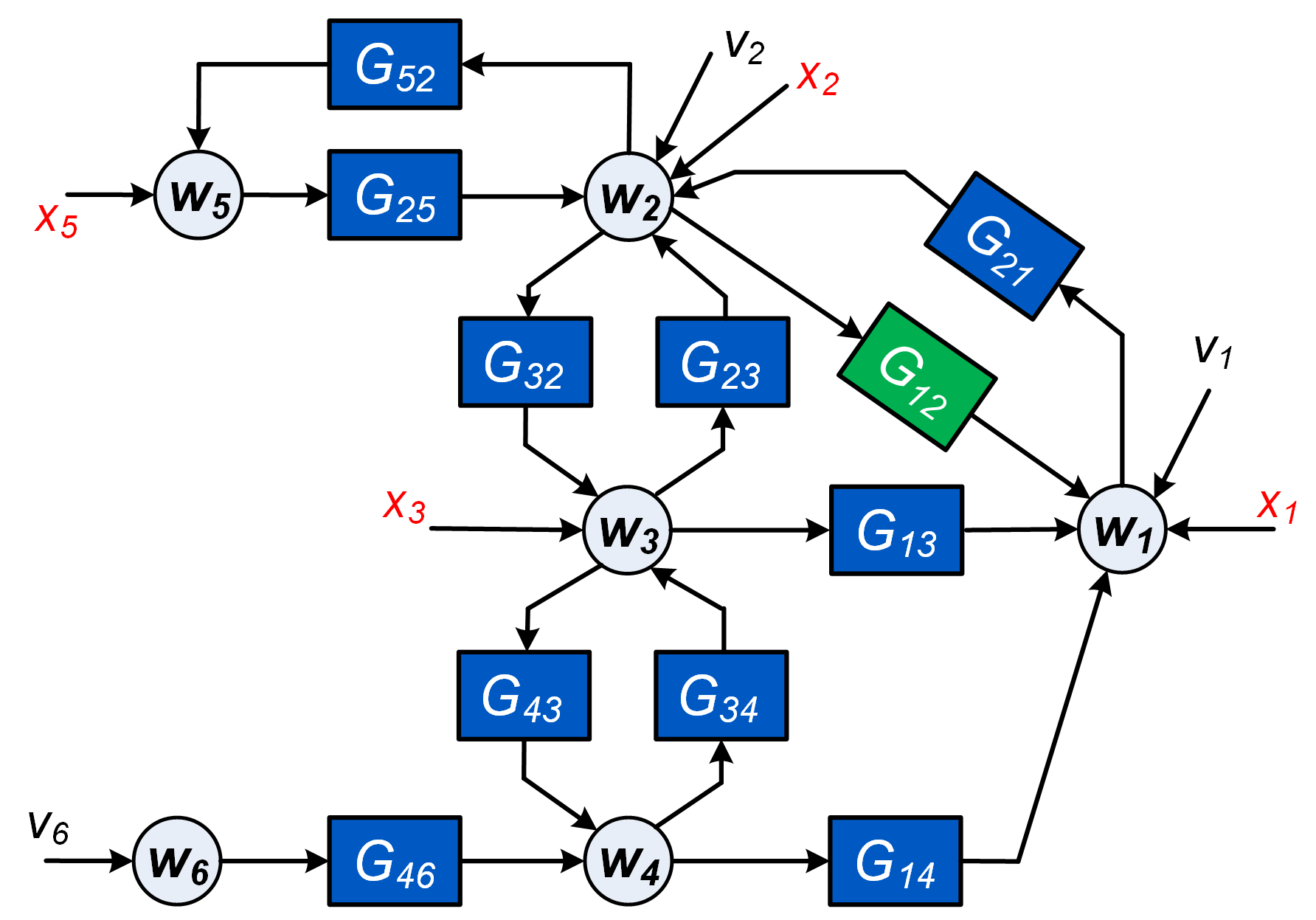}}
	\caption{Six node example with target module $G_{12}(q)$ and correlated disturbances $v_1, v_2$.}
	\label{fig6node}
\end{figure}

We consider the 6-node example with target module $G_{21}$, presented in Section \ref{sec:achiev}, while for ease of reference we repeat its figure as Figure \ref{fig6node}.
Given that the target module is $G_{12}$ we start composing the predictor model with input $w_2$ and output $w_1$. Because of the parallel path connection between $w_2$ and $w_1$ through $w_3$, we add $w_3$ as predictor input. And in order to handle the confounding variable that is present through the correlated disturbances $v_1,v_2$, we add $w_2$ to the predictor output. This leads to a predictor model $(w_2,w_3) \rightarrow (w_1,w_2)$, that satisfies the structural conditions for consistently estimating the target module. With this choice, $w_i = w_2$ and $w_{\smD^c} = w_3$.

When analyzing the data-informativity conditions as developed in this paper, the vector signal composed of $(w_2,w_3)$ would need appropriate excitation. Let us first consider possible locations $(x_5, x_3, x_6)$ for unmeasured disturbance signals. Possible disturbance signals $x_5 = e_5$ and $x_6 = e_6$ will not provide excitation for $(w_2,w_3)$ as these disturbance signals have unmeasured paths to one of the output signals $w_2$ or $w_1$. Therefore their white noise sources are part of $e^{\smY}$, and cannot be used for excitation. A disturbance signal $x_3 = e_3$ will lead to excitation for the input signal $w_3$.

An external excitation signal $x_5 = u_5$, will have an unknown path to output $w_2$, but in the predictor model equation for the target output $w_1$ will not appear as input to an unknown term. As a result $u_5$ will belong to $u^{\perp 1}$ and therefore is available for exciting $w_2$. The conclusion is that two vertex disjoint paths are present between $u_5$ and ($u_3$ or $e_3$) on the one side and $\{w_2,w_3\}$ on the other side, thereby satisfying the conditions for target module data-informativity of $G_{12}$.

When applying the earlier results for full predictor model data-informativity, as presented in \cite{VandenHof&etal_IFAC:23}, the result would be that the present predictor model cannot satisfy the data-informativity conditions by adding a sufficient number of external signals only. Relying on the result of Proposition 5 in \cite{VandenHof&etal_IFAC:23}, we would need to extend the predictor model to $(w_1,w_2,w_3,w_5) \rightarrow (w_1,w_2)$, thus adding $w_5$ as input node signal to be measured, and also adding $w_1$ to the predictor input. This is based on the result of the mentioned Proposition, indicating that each loop in the network around nodes that appear both in the input and the output, should pass through an input node. Sufficient conditions for (full predictor model) data-informativity would then be that $w_5$ and $w_3$ are excited by either a noise signal $(e_3,e_5)$ or an excitation signals $(u_3,u_5)$, and that both $w_1$ and $w_2$ are excited by an external excitations signals $u_1$ and $u_2$. This illustrates how single module data-informativity can lead to significantly more relaxed conditions for instrumenting the network with sensors and actuators.

\section{Conclusions}
It has been shown that data-informativity conditions for MIMO models can be relaxed if we are only interested in identifying a single target module (transfer function) in the MIMO model, while using a direct prediction error method. Only a subset of input signals require excitation, where the subset is constructed on the basis of a disconnecting set property. Starting from an open-loop situation, the results have been formulated for generally interconnected dynamic networks, and graph-based conditions have been developed that guarantee satisfaction of the conditions in a generic sense. The resulting path-based data-informativity conditions are shown to be closely related to earlier derived conditions for generic identifiability of single modules in a dynamic network. This implies that in specified situations, satisfaction of the generic single module identifiability conditions guarantees that the local direct method provides an algorithm for consistently estimating the target module.

\section*{Acknowledegments}
The authors gratefully acknowledge discussions with Xavier Bombois that contributed to the results of this paper.

\appendix

\section{Proof of Proposition \ref{prop3}.}

With the decomposition of input signal $u$ according to (\ref{eq11}), and adding the inputs $u_{\smF}$ to non-parametrized terms, we can write the expression (\ref{equ}) for $u$ by reordering its rows as (leaving out time arguments $(t)$ for brevity):
\beq
\begin{bmatrix} u_i \\ u_{\smD^c} \\ u_{\smE} \\ u_{\smF}\end{bmatrix} =
\begin{bmatrix} 0 & T_{12}(q) & T_{13}(q) & T_{14}(q)\\ T_{21}(q) & T_{22}(q) & T_{23}(q) & T_{24}(q)\\ T_{31}(q) & T_{32}(q) & T_{33}(q) & T_{34}(q) \\ T_{41}(q) & T_{42}(q) & T_{43}(q) & T_{44}(q) \end{bmatrix}
\begin{bmatrix} u_i \\ u_{\smD^c} \\ u_{\smE} \\ u_{\smF} \end{bmatrix} + R_x x.
\eeq
The components $u_{\smF}$ can be removed (immersed) from the description through Gaussian elimination, leading to the expression
\beq
\begin{bmatrix} u_i \\ u_{\smD^c} \\ u_{\smE} \end{bmatrix} =
\begin{bmatrix} 0 & \tilde T_{12}(q) & \tilde T_{13}(q) \\ \tilde T_{21}(q) & \tilde T_{22}(q) & \tilde T_{23}(q) \\ \tilde T_{31}(q) & \tilde T_{32}(q) & \tilde T_{33}(q) \end{bmatrix}
\begin{bmatrix} u_i \\ u_{\smD^c} \\ u_{\smE} \end{bmatrix} + \tilde T_x(q) x.
\eeq
The components of $x$ that appear in the equation for $u_k$, $k \in \{i\cup\mD^c\cup\mT\}$ are then the $x$ components that are directly added to the concerned nodes in the original system representation, plus all those components $x$ that only pass through nodes in $\red{\{}u_{\smF}\red{\}}$ before they reach $\red{\{}u_k\red{\}}$. This is a direct result of Gaussian elimination (immersion) of the nodes $\red{\{}u_{\smF}\red{\}}$ \citep{Dankers&etal_TAC:16}.\\
Because of the disconnecting set property of $\red{\{}u_{\smD^c}\red{\}}$, it follows that $\tilde T_{31}(q) = 0$. As a result, the third block row of the expression can be rewritten as:
\beqr
u_{\smE}(t) & = & (I- \tilde T_{33}(q))^{-1}\tilde T_{32}(q) u_{\smD^c}(t) + \nonumber \\ & & + (I-\tilde T_{33}(q))^{-1}\begin{bmatrix} 0 & 0 & I \end{bmatrix}\tilde T_x(q)x(t).
\eeqr
The components of $x$ that appear in this equation, are the $x$-components that are directly added to $u_{\smE}$, plus the components of $x$ that are directly added to $u_{\smF}$ and only pass through nodes in $\red{\{}u_{\smF}\red{\}}$ before they reach $\red{\{}u_{\smE}\red{\}}$. Differently formulated: the $x$-components that affect the equation for $u_{\smE}$ are those $x$ components that directly enter nodes $\red{\{}u_{\smE\cup\smF}\red{\}}$ and that have a path to $\red{\{}u_{\smE}\red{\}}$ without passing through $\red{\{}u_{i\cup\smD^c}\red{\}}$. Since any path from $u_i$ to $\red{\{}u_{\smE}\red{\}}$ has to pass through $\red{\{}u_{\smD^c}\red{\}}$, it is sufficient to check the passing through $\red{\{}u_{\smD^c}\red{\}}$.
\hfill $\Box$
\section{Proof of Theorem \ref{theo1}.}
We consider the expression $\Delta \bar M_j(q,\theta)\kappa^{[j]}(t)$, where $\bar M_j$ is constructed from $M_j$ by removing the non-parametrized terms, and we decompose the expression as
\beq
\label{eqap1}
\Delta \bar M_j(q,\theta)\kappa^{[j]}(t) = \Delta \bar M_j^u(q,\theta)\kappa_u^{[j]}(t) + \Delta \bar M_j^e(q,\theta)\kappa_e^{[j]}(t).
\eeq
With an appropriate decomposition of $\Delta \bar M_j^u(q,\theta)$, corresponding to (\ref{eq11}) according to:
\beq
  \Delta \bar M_{j}^u(q,\theta) = \begin{bmatrix} \Delta G_{ji}(q,\theta) & \Delta_{j}^a(q,\theta) & \Delta_{j}^b(q,\theta)   \end{bmatrix}
\eeq
as
\beq
 \Delta \bar M_{j}^u(q,\theta)\kappa_u^{[j]} = \begin{bmatrix} \Delta G_{ji}(q,\theta) & \Delta_{j}^a(q,\theta) & \Delta_{j}^b(q,\theta) \end{bmatrix} \begin{bmatrix} u_i \\ u_{\smD^c} \\ u_{\smE} \end{bmatrix}.
\eeq
Since according to (\ref{eqrs}), $u_{\smE}$ can be written as a function of $u_{\smD^c}$ and $x_{\smE^*}$, substituting this in the above expression provides
\beqr
  \begin{bmatrix} \Delta G_{ji}(q,\theta) & \Delta_{j}^a(q,\theta)\! +\!  \Delta_{j}^b(q,\theta) T_s(q) &  \Delta_{j}^b(q,\theta) R_s(q)\end{bmatrix} \nonumber \\ \cdot \begin{bmatrix}  u_i \\ u_{\smD^c} \\ x_{\smE^*} \end{bmatrix}. \nonumber
\eeqr
Denote $\nu := \begin{bmatrix} u_i \\ u_{\smD^c} \end{bmatrix}$. Then we can decompose $\nu$ according to $\nu = \nu_{x_{\smE^*}} + \nu_{\perp x_{\smE^*}}$ and using the fact that
$\Delta_{j}^b(q,\theta)$  is parametrized independently from $\Delta G_{ji}$ and $\Delta_{j}^a$, we can rewrite $\Delta \bar M_{j}^u(q,\theta)\kappa_u^{[j]}$ as
\beqr
 = \begin{bmatrix} \Delta G_{ji}(q,\theta) & \Delta_{j}^c(q,\theta) & | & \Delta_{j}^b(q,\theta) R_s(q)\end{bmatrix}\begin{bmatrix}  \nu_{x_{\smE^*}}\! +\! \nu_{\perp x_{\smE^*}} \\ x_{\smE^*} \end{bmatrix} \nonumber
\eeqr
where $\Delta_{j}^c(q,\theta)$ replaces the term $\Delta_{j}^a(q,\theta) + \Delta_{j}^b(q,\theta) T_s(q)$.
The expression for $\Delta \bar M_{j}^u(q,\theta)\kappa_u^{[j]}$ is now written as a summation of two terms that are mutually uncorrelated: one driven by $x_{\smE^*}$ and $\nu_{x_{\smE^*}}$ and one driven by its orthogonal complement. According to (\ref{eqap1}), for evaluating $\Delta \bar M_j(q,\theta)\kappa^{[j]}(t)$ a third term is added, given by $\Delta \bar M_j^e(q,\theta)\kappa_e^{[j]}$, leading to
\[
 \Delta \bar M_j(q,\theta)\kappa^{[j]}(t)\! =\! \begin{bmatrix} \Delta G_{ji}(q,\theta) & \Delta_{j}^c(q,\theta) &\! |\!\! & \Delta_{j}^b(q,\theta) R_s(q)\end{bmatrix}  \]
\[\hspace*{2cm} \cdot \left( \begin{bmatrix} v_{x_{\smE^*}} \\ x_{\smE^*} \end{bmatrix} + \begin{bmatrix} v_{\perp x_{\smE^*}} \\ 0 \end{bmatrix}\right) + \Delta \bar M_j^e(q,\theta)\kappa_e^{[j]}.
\]
Since both terms $v_{\perp x_{\smE^*}}$ and $\kappa_e^{[j]}$ are orthogonal to $x_{\smE^*}$, it follows now that
$\Eb\|\Delta \bar M_{j}(q,\theta)\kappa^{[j]}(t)\|^2 = 0$ implies $\begin{bmatrix} \Delta G_{ji}(q,\theta) & \Delta_{j}^c(q,\theta) & \Delta \bar M_j^e(q,\theta)\end{bmatrix} =0$ under the condition that $\Phi_{\mu^{[j]}}(\omega) \succ 0$ for almost all $\omega$ with $\mu^{[j]}$ given by
\[
\mu^{[j]} = \begin{bmatrix} \nu_{\perp x_{\smE^*}} \\ \kappa_e^{[j]} \end{bmatrix},
\]
as specified in (\ref{eqmuj}). This implies $\Delta G_{ji}(q,\theta) = 0$ in the minimum of the quadratic cost function. \hfill $\Box$

\section{Proof of Theorem \ref{theox}.}

The proof follows a reasoning that is similar to the proof of Theorem \ref{theo1}. For completeness we include it here.

We consider the expression $\Delta \bar M_j(q,\theta)\kappa^{[j]}(t)$, where $\bar M_j$ is constructed from $M_j$ by removing the non-parametrized terms, and we decompose the expression as
\beqr
\Delta \bar M_j(q,\theta)\kappa^{[j]}(t) & = & \Delta \bar M_j^w(q,\theta)w_{\smD}^{[j]}(t)\! +\!  \Delta \bar M_j^{\xi}(q,\theta)\xi_{\smY}^{[j]}(t) \nonumber \\ & & \ \ \ \ \ \ \  + \Delta \bar M_j^{u}(q,\theta)u^{[j]}(t).
\eeqr
As a consequence of the definition of $w_{\smD^c}$ and the resulting $w_{\smT}$, following Definition \ref{def:disc2}, the term
$\Delta \bar M_{j}^w(q,\theta)w_{\smD}^{[j]}$ can be written as
\beq
 \Delta \bar M_{j}^w(q,\theta)w_{\smD}^{[j]} = \begin{bmatrix} \Delta \bar G_{ji}(q,\theta) & \Delta_{j}^a(q,\theta) & \Delta_{j}^b(q,\theta) \end{bmatrix} \begin{bmatrix} w_i \\ w_{\bar\smD^c} \\ w_{\smE} \end{bmatrix} \nonumber
\eeq
with $\{w_{\bar\smD^c}\} = \{w_{\smD^c}\}\cap \{w_{\smD}^{[j]}\}$. By complementing $w_{\bar\smD^c}$ with the remaining components of $w_{\smD^c}$ and adding $0$ entries to the accompanying $\Delta ^a_j(q,\theta)$, this can be rewritten as
\beq
 \Delta \bar M_{j}^w(q,\theta)w_{\smD}^{[j]} = \begin{bmatrix} \Delta \bar G_{ji}(q,\theta) & \tilde\Delta_{j}^a(q,\theta) & \Delta_{j}^b(q,\theta) \end{bmatrix}\! \begin{bmatrix} w_i \\ w_{\smD^c} \\ w_{\smE} \end{bmatrix}. \nonumber
\eeq
Since according to (\ref{eqrs2}), $w_{\smE}$ can be written as a function of $w_{\smD^c}$ and $x_{\smE^*}$, substituting this in the above expression provides
\beqr
  \begin{bmatrix} \Delta \bar G_{ji}(q,\theta) & \tilde\Delta_{j}^a(q,\theta)\! +\!  \Delta_{j}^b(q,\theta) T_s(q) &  \Delta_{j}^b(q,\theta) R_s(q)\end{bmatrix} \cdot \nonumber \\  \begin{bmatrix}  w_i \\ w_{\smD^c} \\ x_{\smE^*} \end{bmatrix}. \nonumber
\eeqr
%
Denote $\nu := \begin{bmatrix} w_i \\ w_{\smD^c} \end{bmatrix}$ and $\chi:= \{ x_{\smE^*}\cup u^{[j]}\}$ . Then we can decompose $\nu$ according to $\nu = \nu_{\chi} + \nu_{\perp \chi}$ and using the fact that
$\Delta_{j}^b(q,\theta)$  is parametrized independently from $\Delta G_{ji}$ and $\tilde\Delta_{j}^a$, we can rewrite $\Delta \bar M_{j}^w(q,\theta)w_{\smD}^{[j]}$ as
\beq
 = \begin{bmatrix} \Delta G_{ji}(q,\theta) & \Delta_{j}^c(q,\theta) & | & \Delta_{j}^b(q,\theta) R_s(q)\end{bmatrix}\! \cdot\! \begin{bmatrix}  \nu_{\chi} + \nu_{\perp \chi} \\ x_{\smE^*} \end{bmatrix} \nonumber
\eeq
The expression for $\Delta \bar M_{j}^u(q,\theta)w_{\smD}^{[j]}$ is now written as a summation of two terms that are mutually uncorrelated: one driven by $x_{\smE^*}$ and $\nu_{\chi}$ and one driven by a term that is orthogonal to this. For evaluating $\Delta \bar M_j(q,\theta)\kappa^{[j]}(t)$ two additional terms are added, given by $\Delta \bar M_j^{\xi}(q,\theta)\xi_{\smY}^{[j]}$ and $\Delta \bar M_j^{u}(q,\theta)u^{[j]}$.\\
The first of these two terms is possible correlated with $\nu_{\perp\chi}$, while the second is not.\\
It follows now that
$\Eb\|\Delta \bar M_{j}(q,\theta)\kappa^{[j]}(t)\|^2 = 0$ implies $\begin{bmatrix} \Delta \bar G_{ji}(q,\theta) & \Delta_{j}^c(q,\theta)\ |\ \Delta \bar M_j^{\xi}(q,\theta) \end{bmatrix} =0$ under the condition that $\Phi_{\eta^{[j]}}(\omega) \succ 0$ for almost all $\omega$ with $\eta^{[j]}$ given by
\[
\eta^{[j]} = \begin{bmatrix} \nu_{\perp \chi} \\ \xi_{\smY}^{[j]} \end{bmatrix},
\]
as specified in (\ref{eqeta}). This implies that $\Delta \bar G_{ji}(q,\theta) = 0$ in the minimum of the quadratic cost function. \hfill $\Box$

\section{Proof of Proposition \ref{propgraph2}.}

When applying the graph-based results of Proposition \ref{propg} to the results of Theorem \ref{theox}, the graph-based condition becomes to having $dim(\eta^{[j]})$ vertex disjoint paths from all external signals $\{u,e\}$ to $\{\eta^{[j]}\}$. With the expression for $\eta^{[j]}$ (\ref{eqeta}) this becomes: $dim(\eta^{[j]})$ vertex disjoint paths from $\{u,e\}\backslash \{\chi\}$ to $\{w_{i\cup\smD^c},\xi_{\smY}^{[j]}\}$.\\
The white noise signals $\xi_{\smY}^{[j]}$ are typically constructed after spectral factorization of the disturbance process affecting $w_{\smY}$. If in the predictor model, the noise model $H(q,\theta)$ is parametrized fully, this implies that every noise signal $e_{\ell} \in e^{\smY}$, that in the network has a path to a node $w_{\smY}$ with unknown dynamics (i.e. the dynamics in the path is not fixed a priori), will potentially contribute to $\xi_{\smY}^{[j]}$. Note that this does not only refer to noise signals $e$ that have a direct link to $w_{\smY}$ in the network; it also involves noise terms $e$ on unmeasured nodes that are not part of the predictor model. If $H(q,\theta)$ has a structured parametrization, like e.g. a block diagonal form, then we can restrict the class of $e$-signals that should be excluded from the subset of $e_{\smY}$ to those signals $e_{\ell}$ that have a path to an output $w_k \in w_{\smY}$ of which the disturbance component is correlated to the disturbance component of $w_j$.

In the vertex disjoint path condition, this set of signals $e^{\smY}$ is then required to establish the positive definite spectrum condition for $\xi_{\smY}^{[j]}$, which is satisfied by definition. The remaining term then becomes: $dim(\smD^c)+1$ vertex disjoint paths between $\{u,e\}\backslash \{e^{\smY},\chi\}$ and $\{w_{i\cup\smD^c}\}$. Or equivalently $dim(\mD^c)+1$ vertex disjoint paths between $\{u\backslash u^{[j]},e\backslash e^{\smY}\}\backslash \{x_{\smE^*}\}$ and $\{w_{i\cup\smD^c}\}$, which proves the result.
\section{Proof of Corollary \ref{cor10}.}
External excitation signals that are directly entering input nodes $w_{\smD}$ will contribute to $u^{\perp j}$, except when there is a confounding variable between the concerned input and $w_{\smY}$. The input node is then classified to belong to set $\red{\{}w_{\smB}\red{\}}$, \cite{VandenHof&etal_IFAC:23}. For the data-informativity conditions, one can always construct a disconnecting set $\red{\{}w_{\smD^c}\red{\}}$ consisting of nodes that are out-neighbors of $w_i$. For satisfying the structural conditions a. and b., nodes in $\red{\{}w_{\smB}\red{\}}$ are not allowed to be out-neighbors of $w_i$. As a result there always exist a disconnecting set that does not contain nodes in $\red{\{}w_{\smB}\red{\}}$, and external excitation on these nodes will always appear in $u^{\perp j}$.

\bibliographystyle{plain}
\bibliography{Paul_Dynamic_Networks_Library}

\end{document}